\documentclass[twocolumn,trackchanges]{aastex7}

\usepackage{booktabs}
\usepackage{multirow}
\usepackage{amsmath}
\DeclareUnicodeCharacter{2212}{-}

\begin{document}

\title{The GMRT High-Resolution Southern Sky Survey for pulsars and transients - VIII: Orbital Variability and the Evolution of a 1-Day He-WD Millisecond Pulsar J2101--4802}

\author[0009-0002-3211-4865]{Ankita Ghosh}
\affiliation{National Centre For Radio Astrophysics, Tata Institute of Fundamental Research, Pune 411007, India}
\email{ankita@ncra.tifr.res.in}

\author[0000-0002-6287-6900]{Bhaswati Bhattacharyya}
\affiliation{National Centre For Radio Astrophysics, Tata Institute of Fundamental Research, Pune 411007, India}
\email{bhaswati@ncra.tifr.res.in}

\author[0000-0001-6295-2881]{David L.\ Kaplan}
\affiliation{Center for Gravitation, Cosmology, and Astrophysics, Department of Physics, University of Wisconsin-Milwaukee, PO Box 413,
Milwaukee, WI, 53201, USA}
\email{kaplan@uwm.edu}

\author[0000-0002-7833-0275]{David A. Smith}
\affiliation{Laboratoire d’Astrophysique de Bordeaux, Université de Bordeaux, CNRS, B18N, allée Geoffroy Saint-Hilaire, F-33615 Pessac, France}
\email{david.smith@u-bordeaux.fr}

\author[0000-0002-4799-1281]{Andrew Lyne}
\affiliation{Jodrell Bank Centre for Astrophysics, School of Physics and Astronomy, The University of Manchester,
Manchester M13 9PL, UK}
\email{andrew.g.lyne@manchester.ac.uk}

\author[0000-0002-2892-8025]{Jayanta Roy}
\affiliation{National Centre For Radio Astrophysics, Tata Institute of Fundamental Research, Pune 411007, India}
\email{jroy@ncra.tifr.res.in}

\author[0000-0003-0292-4453]{Laila Vleeschower}
\affiliation{Center for Gravitation, Cosmology, and Astrophysics, Department of Physics, University of Wisconsin-Milwaukee, PO Box 413,
Milwaukee, WI, 53201, USA}
\email{vleescho@uwm.edu}

\author[0000-0001-5134-3925]{Gabriella Agazie}
\affiliation{Center for Gravitation, Cosmology, and Astrophysics, Department of Physics, University of Wisconsin-Milwaukee, PO Box 413,
Milwaukee, WI, 53201, USA}
\email{gabriella.agazie@gmail.com}

\author[0000-0002-2554-0674]{Lankeswar Dey}
\affiliation{Department of Physics and Astronomy, West Virginia University, P.O. Box 6315, Morgantown, WV 26506, USA}
\affiliation{Center for Gravitational Waves and Cosmology, West Virginia University, Chestnut Ridge Research Building,
Morgantown, WV 26505, USA}
\email{lanky441@gmail.com}

\author[0000-0002-3764-9204]{Sangita Kumari}
\affiliation{National Centre For Radio Astrophysics, Tata Institute of Fundamental Research, Pune 411007, India}
\email{sangeetakapil8@gmail.com}

\author[0000-0002-2441-4174]{Ujjwal Panda}
\affiliation{National Centre For Radio Astrophysics, Tata Institute of Fundamental Research, Pune 411007, India}
\email{upanda@ncra.tifr.res.in}

\begin{abstract}
We present timing and orbital phase-resolved polarimetry of the millisecond pulsar (MSP) J2101$-$4802, having a spin period of 9.48~ms and dispersion measure (DM) $25.05\ \mathrm{pc\ cm^{-3}}$ discovered with the Giant Meter Radio Telescope (GMRT). From the phase-connected timing of this MSP spanning 3.7 years, we identify that PSR J2101--4802 is in a $\sim$1-day binary orbit with a likely helium–white-dwarf (He-WD) companion having a median companion mass $\simeq0.15\, M_\odot$ (assuming the inclination angle 60\degr), consistent with canonical recycling in the Galactic field. 
The timing solution further reveals an unusually large orbital period derivative, $\dot{P}_b$ ($\sim10^{-11}\,{\rm s\,s}^{-1}$), compared to typical Galactic-field MSP--HeWD binaries, which cannot be explained by the contributions from kinematic effects (Shklovskii and Galactic acceleration) or general-relativistic damping.  Using wideband, full-Stokes observations, we also trace the linear and circular polarization variation across the orbital phase and fit a rotating-vector model (RVM) to its position-angle swing across orbital phase, yielding constraints on the emission geometry (magnetic inclination and impact angle) of this system. The combination of a $\sim$1-day orbit, $\sim0.15\,M_\odot$ companion, modest spin-down power, unusually large $\dot{P}_b$, and phase-locked magnetized intrabinary plasma signatures suggests that PSR~J2101$-$4802 represent a transitional system linking redback-like spiders to detached He--WD MSP binaries.

\end{abstract}


\section{Introduction}\label{sec:intro}

Millisecond pulsars (MSPs) are rotating neutron stars (NSs) with spin periods $P<30$~ms. About 80\% of MSPs are in binaries, typically with white dwarf (WD) companions, but also with main-sequence (MS) stars or other NSs \citep{2008LRR....11....8L,2022ASSL..465....1B}. For the remaining ~20\% of MSPs, no binary companion was reported in the ATNF pulsar catalogue, indicating that these are largely isolated systems originating from binary evolution in which the companion is either completely ablated by the pulsar wind or the binary is disrupted through dynamical interactions \citep{2005AJ....129.1993M,2013IAUS..291..127R}.

According to the theory of pulsar evolution, the more massive star in an initial stellar binary evolves first into an NS; as the companion leaves the main sequence and ascends the red-giant branch, it fills its Roche lobe and transfers matter and angular momentum to the NS, spinning or ``recycling" the NS up to millisecond periods \citep{1982Natur.300..728A,1982CSci...51.1096R}. At the end of this stage, the binary contains an MSP and a stripped stellar core that subsequently becomes either another NS or, more commonly, a WD, depending on its mass \citep{2023pbse.book.....T}. In most systems, the endpoint is an MSP in a binary having an orbital period ranging between a few hours to tens of days, typically in a nearly circular orbit with a low-mass WD companion \citep{2006csxs.book..623T}. Tidal interactions during the Roche Lobe Overflow (RLO) phase circularize the orbit \citep{1992RSPTA.341...39P} and tend to align the pulsar’s spin with the orbital angular-momentum axis \citep{1994ARA&A..32..591P}. For systems that produce helium-core WDs (HeWDs), binary-evolution calculations predict a tight relation between orbital period $P_b$ and WD mass $M_{\rm WD}$ (the TS99 relation; \citealt{1999A&A...350..928T}, refined by more recent studies; e.g. \citealt{2014A&A...571A..45I,2016A&A...595A..35I}), which generally holds for Galactic-field MSP binaries \citep[e.g.,][]{2018ApJ...864...30H}. 
Another notable subset of recycled MSPs is the ``spiders'', with compact orbits and irradiated, low-mass companion stars. In these systems, the pulsar’s energetic wind ablates the companion, producing radio eclipses and ellipsoidal modulation from the tidal distortion of the companion
star. These systems are commonly divided into \emph{black widows}, with companion masses, $M_c\!\lesssim\!0.05\,M_\odot$, and \emph{redbacks} (non-degenerate, nearly Roche-lobe–filling companions) with $M_c\!\sim\!0.1$–$0.7\,M_\odot$ \citep{2013IAUS..291..127R}. The intrabinary shocks and variable mass loss observed in these systems provide a comprehensive view of their emission mechanisms and the evolutionary pathways of these systems \citep{2019Galax...7...93H,2020ApJ...904...91V}.

Precise timing of MSPs can reveal diverse binary classes (e.g.\ black widows/redbacks, MSP--WD systems, hierarchical triples), enabling mass measurements of the companion star as well as the NS and tests of strong-field gravity \citep{2016arXiv160501665A}. It can provide constraints on different binary evolution paths and Galactic dynamics. It is foundational for pulsar timing arrays (PTAs), where long-term timing of many exceptionally stable MSPs is used to detect nanohertz gravitational waves and to probe supermassive black-hole binaries and stochastic backgrounds \citep[e.g.,][]{2016MNRAS.458.1267V}. Polarization study of MSPs can constrain their emission geometry and magnetospheric behavior through rotation measures (RMs) and polarization-angle (PA) swings \citep{1969ApL.....3..225R}.

 In recycled MSP--WD binaries with day-long, nearly circular orbits, the orbital period derivative is expected to be dominated by kinematic effects, primarily the Shklovskii contribution and acceleration in the Galactic field, while intrinsic relativistic effects such as gravitational-wave damping are typically negligible \citep[e.g.,][]{1991ApJ...366..501D}. In some MSP binaries with low-mass companions, large orbital period derivatives ($\dot{P}_b$) have been linked to changes in the companion’s gravitational quadrupole moment driven by magnetic activity and tidal dissipation \citep[e.g.,][]{1994ApJ...436..312A,2001A&A...379..579D,2011MNRAS.414.3134L}.
An additional contribution to an apparently excess $\dot{P}_b$ can also arise from acceleration of the inner binary due to a third body in a hierarchical system, which Doppler-shifts the observed orbital period and can mimic secular derivatives over a limited timing baseline \citep[e.g.,][]{1994ApJ...432..239M,1996ASPC..105..525A,2025A&A...697A.166D}. Apparent excess orbital-period derivatives can also arise from timing-model covariances and systematics (e.g.\ between $P_b$, $\dot{P}_b$, and astrometric parameters), especially when orbital-phase coverage is non-uniform, or the timing baseline is limited \citep[e.g.,][]{2016MNRAS.458.3341D}.
 Motivated by these considerations, precise measurements of $\dot{P}_b$ in otherwise canonical MSP--WD binaries provide a sensitive probe of both astrophysical effects and limitations in current timing models. In this work, we present a detailed timing analysis of PSR~J2101$-$4802, focusing on the origin and significance of its measured orbital period derivative and assessing whether the observed excess can be explained by known kinematic effects.

PSR~J2101$-$4802 is a 9.48-ms, Galactic-disk MSP with a DM of $25.05\,\mathrm{pc\,cm^{-3}}$. It was discovered in the GMRT High-Resolution Southern Sky Survey \citep[GHRSS\footnote{\url{http://www.ncra.tifr.res.in/~bhaswati/GHRSS.html}},][]{2016ApJ...817..130B, 2019ApJ...881...59B} by Bhattacharyya et~al.\ (in preparation). 
For PSR~J2101$-$4802, interferometric continuum imaging followed by multiple phased-array (PA) beamforming localized the pulsar to $\sim1''$ \citep{2023ApJ...947...88S}, facilitating rapid phase connection and enabling robust measurements of binary parameters. Using the flux density measurements for PSR~J2101$-$4802 reported in \citet{2023ApJ...947...88S} --- $S_{400} = 10.2 \pm 0.6$~mJy, $S_{650} = 4.0 \pm 0.1$~mJy, and $S_{1280} = 1.23 \pm 0.09$~mJy --- gives an average spectral index of $\alpha \approx -1.8$, consistent with the steep spectra typical of MSPs. However, \citet{2024ApJ...971..174S} reported significant epoch-to-epoch spectral variability, including evidence for spectral turnover in several individual epochs.

In this paper, we present timing, astrometric, and polarimetric observations of PSR~J2101$-$4802. We place the system in the context of MSP--HeWD binary evolution, discuss the implications for its kinematics, and highlight how its polarization characteristics inform the emission geometry and magnetospheric physics.

In Section~\ref{sec:observation}, we describe the upgraded Giant Metrewave Radio Telescope (uGMRT) and Parkes observations of PSR~J2101$-$4802. Section~\ref{sec:Timing} presents the phase-connected timing analysis, while Section~\ref{sec:kinematics} discusses the kinematic corrections applied to the measured spin-period derivative ($\dot{P}$) and $\dot{P}_b$. Wideband timing of the Parkes data is presented in Section~\ref{sec:wideband-timing}, and the frequency-dependent evolution of the pulse profile is characterized in Section~\ref{sec:Profile evolution}. The polarization properties and emission geometry of the MSP are discussed in Sections~\ref{subsec:polarization} and~\ref{sec:emission geometry}, respectively. Finally, we discuss our results in Section~\ref{sec:Discussion} and summarize our conclusions in Section~\ref{sec:Summary}.

\section{Observations and data analysis} \label{sec:observation}

\renewcommand{\tabcolsep}{0.5pt}
\begin{table}[!htb]
\begin{center}
\caption{Summary of observations}
\footnotesize{
\label{tab:1}
\begin{tabular}{cccccccc}
\toprule
Receiver  &  Frequency & Mode$^{a}$ & T$_{\rm res}$$^{b}$ &  N$_{\rm ch}$$^{c}$ & $\sigma_{\rm ToA}$$^{d}$ & No. of  &  Span \\ 
band & (MHz) &  & ($\mu$s) & &($\mu$s)  &epochs & \\
\midrule
 \multirow{2}{*}{GMRT Band 3}  & \multirow{2}{*}{400} & IA & 81.92 & 4096 & 5.36 & 7 & 2019$-$2022 \\
                         &                             & PA & 81.92 & 4096 & 1.03 & 45 & 2022$-$2025 \\
 \midrule
GMRT Band 4  & 650 & PA & 81.92 & 4096  & 2.72 & 4 & 2023  \\
 \midrule
GMRT Band 5$^{\dagger}$ & 1280 & PA & 81.92 & 4096 & 5.72 & 1 & 2023-2025 \\
 \midrule
Parkes UWL & 2368 & CD & 9.25 & 3328 & 2.03 & 7 & 2024--2025 \\
\bottomrule
\end{tabular}
}
\end{center}
{\footnotesize {\bf{Notes.}}\\
$^{a}$IA: Incoherent Array observations. PA: Phased Array observations. CD: Observations with coherent de-dispersion \\
$^{b}$Time resolution.\\
$^{c}$Number of channels.\\
$^{d}$TOAs are calculated for each mode of observation of an average duration of 45 minutes in band 3, band 4, and band 5. The mentioned $\sigma_{\rm ToA}$ for each observing band is the mean value.\\
$^{\dagger}$Band 5 is used only for profile study due to poor signal-to-noise ratio.}
\end{table}

The GMRT is a radio interferometer, operating across five bands from $\sim150$ to $1450$~MHz \citep{1991ASPC...19..376S, 2017CSci..113..707G}. 
We observed PSR~J2101$-$4802 with the uGMRT  \citep{1991ASPC...19..376S, 2017CSci..113..707G, http://dx.doi.org/10.1142/S2251171716410117} in band$-$3 (300$−$500~MHz), band$-$4 (550$−$750~MHz), and band$-$5 (1050$-$1450~MHz) in several epochs as listed in Table \ref{tab:1}. The data were incoherently dedispersed at a central frequency of 400~MHz and 650~MHz with a bandwidth of 200~MHz at DM of 25.055~pc~cm$^{-3}$. We recorded Stokes-I data at a rate of 48~MB/s, employing 8-bit sampling, and used 4096 channels with a sampling time of 81.92~$\mu$s.
To effectively mitigate the impact of narrowband and short-duration broadband radio frequency interference (RFI), we used a real-time RFI filter \citep{2019JAI.....840006B,2022JAI....1150008B} as well as RFI mitigation software in conjunction with the GMRT pulsar tool (gptool\footnote{\url{https://github.com/chowdhuryaditya/gptool}}).
Then, to obtain the folded profile from each epoch’s observations, we used the {\texttt{PRESTO}} \citep{2002AJ....124.1788R} task {\texttt{prepfold}} to perform incoherent dedispersion on the filterbank data and fold each data set at the pulsar's periodicity, producing folded profiles of 30-minute integration. The band--3 incoherent-array (IA) data were used only while obtaining the initial timing solution, and were not included in the final phase-coherent analysis. Similarly, the band--5 observations were used solely to investigate pulse profile evolution with frequency and were not incorporated into the timing analysis due to their lower sensitivity.\par

We observed PSR~J2101$-$4802 using the Parkes Ultra-Wideband (UWL) receiver \citep{2020PASA...37...12H} at a frequency spanning 704 to 4032~MHz, divided into 3328 channels with a frequency resolution of 1~MHz. The data were coherently de-dispersed and folded using the pulsar's timing solution obtained from the previous GMRT observations. The folded data were organized into subintegrations with a duration of 30 seconds and 1024 phase bins per pulsar period. Table \ref{tab:1} provides a summary of these observations. Before observing the target source with Parkes UWL, we briefly monitored a pulsed calibration signal injected into the low-noise amplifiers. We used the pipeline {\texttt{PSRPYPE}}\footnote {\url{https://github.com/vivekvenkris/psrpype/}} for RFI mitigation. The remaining RFI was manually mitigated in the time and frequency domains. {\texttt{PSRCHIVE}} \citep{2004PASA...21..302H} was used for the data processing. The routine {\texttt{pac}} corrects for instrumental gain and phase differences using the pulsed calibrator, carries out flux calibration obtained from observations of PKS~B1934$–$68, and corrects for instrumental leakage terms using the Polarimetric Calibration Model (PCM; \citealt{2004ApJS..152..129V}). To improve the signal-to-noise ratio (SNR), we averaged across frequency and phase bins by factors of 8. This results in a calibrated folded profile comprising 4 Stokes parameters, 416 frequency channels, and 128 phase bins for each epoch.\par
Consecutive observations were combined using the task {\texttt{psradd}}, generating a single output file. A time drift observed in the combined file was rectified by updating the pulsar ephemeris. We used this combined average profile to obtain the average polarization fraction and rotational measure of the pulsar.

\subsection{Timing analysis of J2101--4802} \label{sec:Timing}

Pulsar timing involves an iterative process of fitting a model based on trial values of rotational, astrometric, and binary parameters to the times of arrival (ToAs), finally yielding a timing solution \citep{2004hpa..book.....L}.\par

We produced the ToAs for all data sets using the standard template matching technique employed in pulsar timing. We created separate templates for each observing setup, uGMRT band--3, uGMRT band--4, and Parkes observations, each constructed from the corresponding integrated profile for that band. For frequency-resolved ToAs, we created a frequency-resolved standard profile by iteratively running \texttt{paas} on the integrated profile. The ToAs were obtained via the {\texttt{pat}} command.
The significant decrease in intensity at frequencies above 2000~MHz in the Parkes observations results in large ToA uncertainties in these bands. For the timing analysis, we therefore discarded the ToAs above 2000~MHz.\par

We use the pulsar position measured from uGMRT imaging by \citep{2023ApJ...947...88S} and adopt their reported coordinates as the reference position for our timing analysis. We conduct a least-squares fitting process for the pulsar parameters by minimizing the sum of the squared timing residuals, defined as the differences between actual ToAs and the model predictions. As the orbit of PSR~J2101$-$4802 shows drifts in the pulse phase in the folded data due to unmodelled binary motion, we assumed it to be a compact binary system. We estimated the variation of the observed spin period by fitting a period to the ToAs for each observing epoch. Then we used fitorbit\footnote{\url{https://github.com/vivekvenkris/fitorbit}} to model the spin-period variations across multiple epochs, fitting for binary and astrometric parameters to obtain an initial timing solution.\par

The parameters from this incoherent solution were used as input parameters for obtaining a phase-coherent solution for the system using the pulsar timing package {\texttt{TEMPO2}} \citep{2006MNRAS.369..655H} as well as {\texttt{PINT}} \citep{2018AAS...23145309L, 2021ApJ...911...45L}. The phase connection step is where the precise rotation numbers between the arrival times are definitively determined, leading to the refinement of the timing solution through the optimization of pulsar parameters. ToAs for both the uGMRT and Parkes data were generated iteratively. We first folded the data using the initial timing solution obtained from {\texttt{fitorbit}}, generated preliminary ToAs, and derived a first phase-connected solution. This solution was then used to re-fold and re-align the data in a consistent phase convention, after which the ToAs were regenerated and the timing fit repeated.\par

We performed the timing analysis using the ELL1 binary model \citep{2001MNRAS.326..274L} implemented in {\texttt{TEMPO2}}, which is appropriate for low-eccentricity systems where the longitude of periastron is poorly defined, as is the case for PSR~J2101$-$4802. The derived phase-connected timing model includes the intrinsic spin and spin-down parameters, astrometric parameters (position and proper motion), and five Keplerian orbital parameters together with the orbital period derivative $\dot{P}_b$.  Using the resulting parameter file, we refolded the data from each observing epoch, which removed residual pulse-phase drifts due to previously unmodelled binary motion and improved the overall timing precision.\par

Time-variable DM was included in the model via a piecewise-constant model (“DMX”) to measure DM variations between subsequent observations to produce a DM time series. DMX is a method developed by the NANOGrav collaboration \citep[e.g.][]{2017ApJ...841..125J, 2021ApJS..252....5A}. We tested simpler dispersion models using only a single DM term, and then adding DM1 provides only a modest improvement and still leaves a noticeably worse fit than the DMX case. In contrast, the DMX model reduces the rms from 5.03~$\mu\mathrm{s}$ to 3.50~$\mu\mathrm{s}$ and reduced $\chi^2$ from 1.68 to 1.00. We adopt 15-day DMX bins, chosen by comparing fits with several bin sizes using the Akaike Information Criterion (AIC); 15-day bins provide the best balance between reducing $\chi^2$ and avoiding overfitting while ensuring at least two ToAs per bin, derived from independent frequency sub-bands. The mean DM over the data span is $\sim 25.051$~pc~cm$^{-3}$, with epoch-to-epoch variations described by the DMX terms.

\section{Results} \label{Results}

This 9.48-ms MSP is found to be in a 1-day orbit with a low-mass companion. For an assumed pulsar mass of 1.4~M$_\odot$, the companion mass is 0.13~M$_\odot$ for an orbital inclination of $90^{\circ}$, with a median value of 0.15~M$_\odot$ for an inclination of $60^{\circ}$. The near-day-long orbital period made uniform orbital-phase coverage challenging, particularly for the uGMRT, where PSR~J2101$-$4802 is observable for only $\sim$4\,hr per day. As a result, repeated observations often sample the same, relatively small portion of the orbit, which can introduce parameter covariances (e.g., between $P_b$ and $\dot{P}_b$) and complicate precise measurements of orbital characteristics. Unlike many similar MSP--WD binaries with low-mass companions in nearly circular orbits, where no statistically significant orbital period derivative is detected \citep[e.g.][]{2005ApJ...620..405S, 2005ASPC..328..373J, 2021ApJS..252....5A}, we find a significant measurement of the orbital period derivative in PSR~J2101$-$4802, with $\dot{P}_{\rm b} \simeq -1.2(2)\times 10^{-11}$.\par
\begin{figure*}[ht!]
\begin{center}
\includegraphics[width=\textwidth,angle=0]{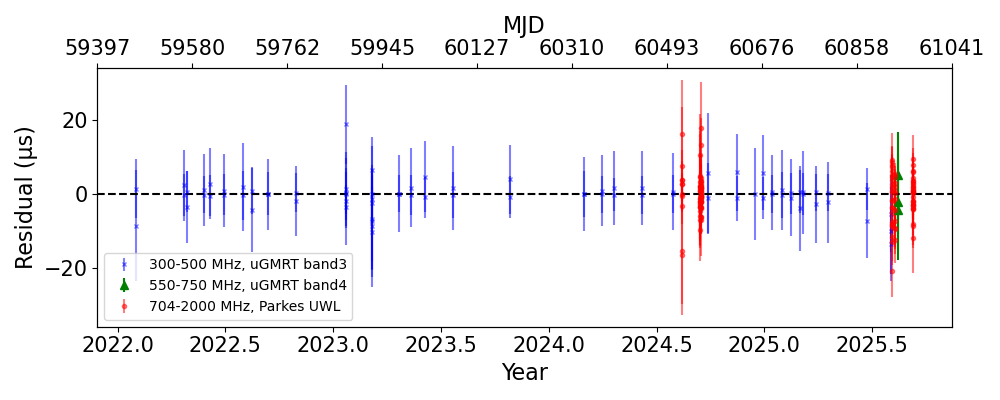}
\caption{Post-fit timing residual of J2101--4802 using uGMRT and Parkes observations showing 3.7 years of timing data.}
\label{fig:timing1}
\end{center}
\end{figure*}

\begin{figure*}[ht!]
\begin{center}
\includegraphics[width=\textwidth,angle=0]{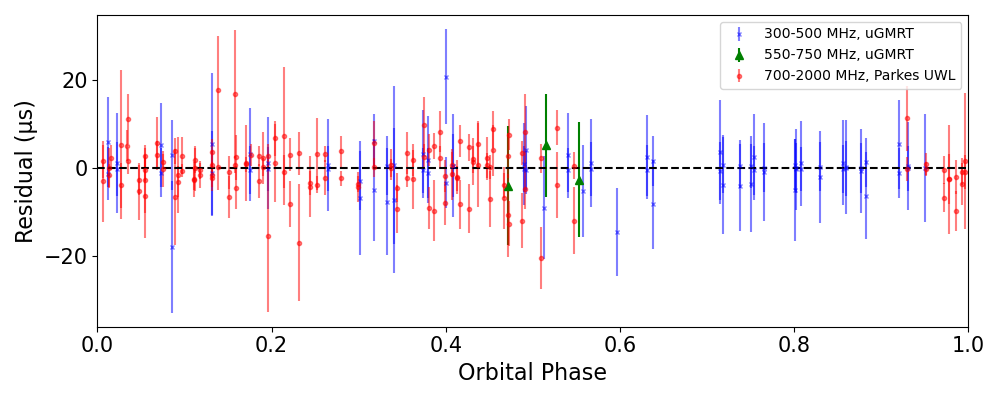}
\caption{Post-fit timing residuals of J2101--4802 using uGMRT and Parkes observations as a function of orbital phase. Since the system is nearly circular, we define the orbital phase relative to the orbital period ($P_b$), with phase 0 corresponding to the time of ascending node ($T_{\rm ASC}$).}
\label{fig:timing2}
\end{center}
\end{figure*}

 The resulting timing model after including DMX terms is presented in Table~\ref{tab:2} and yields an rms residual of 3.50~$\mu$s and a $\chi^2=171$. The timing residuals are shown in Figure~\ref{fig:timing1} and Figure~\ref{fig:timing2} as a function of time and orbital phase.\par

\subsection{Possible systematics from near 1--day orbital period}\label{sec:systematics}

The orbital period, $P_b = 0.99924$~days, differs from Earth's sidereal day ($P_\mathrm{sid} = 0.99727$~days) by only $\Delta P_b = 0.00197$~days ($\approx 2.83$~min). This, in principle, could cause daily observations to repeatedly sample similar orbital phases and introduce covariances among $P_b$, $\dot{P}_b$, and the astrometric parameters. In this section, we put forward the arguments that lead us to believe the detected $\dot{P}_b$ is intrinsic in nature.
\begin{enumerate}
    \item The majority of the phase bins contain $\geq$10 ToAs (except for a modest deficit in the phase range 0.6--0.7) obtained from Parkes UWL and uGMRT observations.

    \item These telescopes observe PSR J2101--4802 at different local sidereal times, sampling different orbital phases. Moreover, the $\dot{P_b}$ detection is consistent in both the uGMRT and Parkes datasets individually as well, arguing against systematics coming from a single telescope.

    \item The offset of $P_b$ from the sidereal day defines a time duration over which the our observations drift through the complete range of orbital phases, sampling the full orbit of the MSP (i.e., beat period),
\begin{equation}
    P_\mathrm{beat}
        = \frac{P_b}{\Delta P_b}
        \approx 506.3~\mathrm{days}
        = 1.4~\mathrm{yr}.
    \label{eq:beat}
\end{equation}
Thus, over the 3.7-year timing baseline, the observations therefore drifted through full orbital phase space for $(3.7/P_\mathrm{beat})=  2.7$ times.  Any systematics tied to Earth's rotation would be modulated at the beat period and would average down over multiple cycles, rather than accumulating coherently as a secular $\dot{P}_b$. Since our observations sampled the full orbital phase nearly three times, it is possible that the observed $\dot{P}_b$ is likely intrinsic to the system.

    \item In addition, we assessed the significance of including the orbital period derivative by comparing nested timing models with and without $\dot{P}_{\rm b}$ before applying DMX terms. The fit improves from $\chi^2=369$ (209 degrees of freedom) to $\chi^2=352$ (208 degrees of freedom) when $\dot{P}_{\rm b}$ is included. An F-test yields a probability of $p\simeq1.6\times10^{-3}$ implying that introducing $\dot{P}_{\rm b}$ leads to a statistically significant improvement in the fit. Also, we use information-criterion comparisons, AIC (Akaike Information Criterion) and BIC (Bayesian Information Criterion), which are model-selection statistics that balance goodness-of-fit against model complexity, with lower values indicating a statistically preferred model after accounting for the number of fitted parameters. Consistently, information-criterion comparisons ( $\Delta{\rm AIC}\approx-15$ and $\Delta{\rm BIC}\approx-12$) strongly favour the model including $\dot{P}_{\rm b}$.

    \item To further validate the robustness of the $\dot{P}_b$ detection against the non-uniform orbital phase sampling caused by the nearly 1-day orbital period, we performed two sets of simulations, both with a true $\dot{P}_b=0$, to see if either would recover a non-zero $\dot{P}_b$. In the first, we generated $\sim$200 realizations using the actual ToA timestamps (real sampling), while in the second, we used randomly distributed timestamps over the same 3.7-year baseline (random sampling). Fake ToAs were generated from the best-fit timing model but with $\dot{P}_b = 0$, and all timing parameters were subsequently fitted freely in each realization. The noise assigned to each simulated ToA was randomly drawn from the Gaussian distribution of the measured ToA uncertainties. In the real-sampling case, only the noise realization varied between simulations, whereas in the random-sampling case, both the observing timestamps and the noise were regenerated for each realization. In both cases, $\sigma_\mathrm{sim}$ from the real sampling is in agreement with the $1\sigma$ uncertainty on $\dot{P}_b$ from the timing fit ($\sim 10^{-12}$~s~s$^{-1}$), confirming that the simulations reproduced the noise properties and parameter covariances of the real dataset.  Notably, though, in both cases the recovered  $\dot{P}_b$ distributions had mean values of order $10^{-14}$\,s\,s$^{-1}$, much less than the $\dot{P}_b$ we measure. These simulations demonstrate that the measured $\dot{P}_b$ is unlikely to be a consequence of the sampling pattern, lending further confidence to its intrinsic origin.

\end{enumerate}
 
Additionally, to check the possible detection of higher order orbital period derivatives, we have re-fitted the timing model using the \textsc{BTX}
binary framework, which parametrises the orbital evolution
through the orbital-frequency derivatives FB0, FB1, and
FB2, thereby allowing a direct measurement of the second
orbital-period derivative $\ddot{P}_b$. This resulted in insignificant FB2 detection, and the fit does not improve (an F-test yields $p = 0.51$).

\begin{table*}
\begin{center}
{\footnotesize
\caption{Timing parameters for PSR J2101$-$4802.
\label{tab:2}}
\begin{tabular}{lc}
\hline
Parameters & Values  \\
\hline
Timing Data Span     \dotfill     & 59611.3$-$60927.7 \\
Period epoch (MJD)\dotfill          &   60269.50 \\
Total time span (year) \dotfill       & 3.70   \\
Number of TOAs\dotfill  & 216  \\
Post-fit residual rms ($\mu$s)\dotfill       & 3.50    \\
Chi-square \dotfill       & 171 \\
Reduced Chi-square\dotfill       & 1.0 \\
\hline
Right ascension (J2000)\dotfill &
21$^\mathrm{h}$01$^\mathrm{m}$55\fs0463(1)  \\
Declination (J2000)\dotfill     &
$-$48\degr01\arcmin59\farcs136(1)  \\
Spin frequency $f$ (Hz)\dotfill & 105.479942441562(3) \\
Spin frequency derivative $\mathrm{\dot{f}}$ (Hz s$\mathrm{^{-1}}$)\dotfill & $-$1.782(5)$\times$10$^{-16}$  \\
Dispersion measure $\mbox{DM}$ (pc~cm$^{-3}$) &  25.05086(6)      \\
Proper motion in R.A.  $\mbox{DM}$ (mas~yr$^{-1}$) & 12.7(1.9) \\
Proper motion in Decl.  $\mbox{DM}$ (mas~yr$^{-1}$) & -4.4(3.5) \\
Orbital period $\mathrm{P_{b}}$ (days)\dotfill &  0.999238195(2)  \\
orbital period 1st derivative, $\mathrm{\dot{P}_{\rm b}}$  ($\mathrm{s\, s^{-1}}$) \dotfill                             & 1.2(2) $\times 10^{-11}$  \\
Projected semi-major axis $\mathrm{x}$ (lt-s) &   0.9585359(1)  \\
Epoch of ascending node $\mathrm{T_{ASC}}$ (MJD) \dotfill &  60269.7303444 (7) \\
$\epsilon_{1} (e\cos\omega)$$^\ast$  & 8(16) $\times 10^{-7}$ \\

$\epsilon_{2} (e\sin\omega)$$^\ast$  & 1.4(1.1) $\times 10^{-6}$ \\

Planetary Ephemeris                & DE200 \\
Time units                         & TDB \\
Flux density at 400 MHz, $S_{400}$ $\mbox{(mJy)}$     & 10.2(6) \\
 \hline
 \multicolumn{2}{c}{Derived parameters} \\
  \hline
Galactic longitude, $\mathrm{l}$ (\degr) \dotfill    & 351.485  \\
Galactic latitude, $\mathrm{b}$ (\degr) \dotfill    & -41.289   \\
DM distance (kpc) \dotfill  & 0.992 $^\dagger$       \\
 & 2.13$^\ddagger$ \\
Mass function $\mathrm{M_{\odot}}$\dotfill &     0.000947041(2) \\
Companion mass$^{**}$, $\mathrm{M_c}$  (M$_{\odot}$)      \dotfill   &   0.1274 $|$ 0.1485 $|$ 0.3167  \\ 
Observed spin period derivative, $\mathrm{\dot{P}_{\rm obs}}$ ($\mathrm{s\, s^{-1}}$) \dotfill                             & 1.596(1) $\times 10^{-20}$  \\
Energy loss rate $\mathrm{\dot{E}}$ ($10^{32} \rm \, erg \, s^{-1}$)\dotfill   &  6       \\
Characteristic age, $\mathrm{\tau_c}$ (Gyear)\dotfill  &  11.6          \\
Surface magnetic field, $\mathrm{B_0}$ (Gauss)\dotfill      &  $3.6 \times 10^{8}$        \\ \hline
\end{tabular}
}
\end{center}

{\footnotesize {\bf{Notes.}}\\
We note that the calculated DM distance is model-dependent.
The numbers in the parentheses are uncertainties in the preceding digits, and errors correspond to 1$\sigma$.\\
 $^\ast$ $e$ is the orbital eccentricity, and $\omega$ is the longitude of periastron (the angle from the ascending node to periastron, measured in the orbital plane).\\
$^\dagger$ using the NE2001 electron density model \citep{2002astro.ph..7156C}.\\
$^\ddagger$ using the YMW16 electron density model \citep{2017ApJ...835...29Y}. \\
 $^{**}$ The quoted companion mass range was obtained from the mass function, assuming a neutron star mass of $1.4\,M_\odot$  and orbital inclinations of $i=90^\circ$ (for minimum companion mass), $i=60^\circ$ (for median companion mass), and $i=26^\circ$ (upper limit corresponding to the 90\% confidence level for a random distribution of inclination angle)\\
The parameter file can also be found \href{https://drive.google.com/file/d/1OmKjgYGyQB5HqJciosGui8ydMSLNSND2/view?usp=sharing}{here}}.
\end{table*}

\subsection{Kinematic corrections to $\dot{P}$ and $\dot{P}_b$}\label{sec:kinematics}

The observed spin-period derivative, $\dot{P}_{\rm obs}$, includes kinematic contributions from both the Shklovskii effect and acceleration in the Galactic gravitational potential, such that
$\dot{P}_{\rm obs} = \dot{P}^{\rm int} + \dot{P}^{\rm Shk} + \dot{P}^{\rm Gal}$.
The DM distance inferred from the YMW16 and NE2001 electron-density models \citep{2002astro.ph..7156C,2017ApJ...835...29Y} are $d = 1 - 2$~kpc respectively.
Therefore, using the measured proper motion $(\mu_\alpha,\mu_\delta) = (12.7,-4.4)\,{\rm mas\,yr^{-1}}$, corresponding to a total proper motion $\mu \simeq 13\,{\rm mas\,yr^{-1}}$, a representative DM distance of $d = 1$~kpc (taking YMW16 model), and a spin period $P = 9.48$~ms, we obtain a Shklovskii contribution \citep{1970SvA....13..562S}
$\dot{P}^{\rm Shk} = \frac{P\mu^2 d}{c} \simeq (4.0 \pm 1.4)\times 10^{-21}$,
where the quoted uncertainty reflects only the uncertainty in the proper motion. The uncertainty associated with the DM-based distance is $\sim$20–30\%  and spans $\sim$1–2~kpc when comparing the YMW16 and NE2001 electron-density models and, therefore, would dominate the total error but is not included here.\par
The Galactic acceleration term, $\dot{P}^{\rm Gal}$, is evaluated independently following \citet{1991ApJ...366..501D}, adopting $K_z$ (the vertical component of the Galactic gravitational acceleration (perpendicular to the Galactic plane) at the pulsar’s location above the plane) from \citet{2004MNRAS.352..440H} and standard Galactic constants. Using the pulsar’s Galactic coordinates $(l,b)=(351\fdg485,-41\fdg289)$ and $d=1$~kpc yields
$\dot{P}^{\rm Gal} \simeq (-1.0 \pm 0.2)\times10^{-21}.$
Thus, the total kinematic contribution is
$\dot{P}^{\rm D} = \dot{P}^{\rm Shk} + \dot{P}^{\rm Gal} \simeq (3.0 \pm 1.4)\times10^{-21}$, implying an intrinsic spin-down rate of $\dot{P}_{\rm int} \approx 1.3\times10^{-20}$.
After correcting for kinematic effects, the intrinsic spin-down implies a spin-down luminosity $\dot{E} \simeq 6.0 \times 10^{32}\ {\rm erg,s^{-1}}$,
where $\dot{E} \equiv 4\pi^2 I \dot{P} P^{-3}$ and we assume a neutron-star moment of inertia $I = 10^{45}$~g~cm$^2$ \citep{2004hpa..book.....L}. The characteristic age is $\tau_c \equiv P/(2\dot{P}) \sim 11.6$~Gyr, and the inferred surface dipole magnetic field strength is $B_0 \equiv 3.2\times10^{19}(P\dot{P})^{1/2} \simeq 3.6\times10^8$~G, consistent with typical values for recycled MSPs.

The DM-based distance range $d\simeq1$–$2$~kpc yields a transverse velocity
$v_T = 4.74\,\mu d \simeq 60\text{--}120\ \mathrm{km\ s^{-1}}$,
consistent with the Galactic-disk MSP population. The timing parallax is $\tau_{\rm PX} = 900 \,\mathrm{ns}(d/1 \mathrm{kpc})$ towards this pulsar. Given that the achieved rms timing residual of MSP J2101--4802 is $\sim4~\mu$s, the timing parallax cannot be usefully constrained in the present data set.\par

The observed orbital-period derivative can also be written as \citep{2009MNRAS.400..805L,2012MNRAS.423.3328F}
\begin{equation}
\dot{P}_b=\dot{P}_b^{\dot{m}}+\dot{P}_b^{T}+\dot{P}_b^{D}+\dot{P}_b^{\rm GW},
\end{equation}
with $\dot{P}_b = 1.2(2)\times10^{-11}$ for PSR~J2101$-$4802. For a galactic He–WD companion, both mass-loss ($\dot{P}_b^{\dot{m}}$) and tidal ($\dot{P}_b^{T}$) contributions are expected to be negligible and are therefore ignored here \citep{2012MNRAS.423.3328F, 2013Sci...340..448A}.

The Doppler term can be written as $\dot{P}_b^{D}=\dot{P}_b^{\rm Gal}+\dot{P}_b^{\rm Shk}$. Using $d=1$~kpc and $P_{\rm b}=0.999$~d, we find a Galactic contribution
$\dot{P}_b^{\rm Gal}=(-9.4\pm2.0)\times10^{-15}$,
and a Shklovskii contribution
$
\dot{P}_b^{\rm Shk}=\frac{\mu^2 d}{c}P_b=(3.6\pm1.3)\times10^{-14},
$
yielding
$
\dot{P}_b^{D}=(2.7\pm1.3)\times10^{-14}.
$

For a circular binary in general relativity, assuming $m_p=1.4\,M_\odot$, $m_c=0.15\,M_\odot$, and $P_b=0.999$~d, the predicted gravitational-wave contribution is negligible,
$
\dot{P}_b^{\rm GW}=-9.8\times10^{-17}.
$

Taken together, the Doppler and GR contributions,
$
\dot{P}_b^{\rm D}+\dot{P}_b^{\rm GW}\simeq(2.7\pm1.3)\times10^{-14},
$
is more than two orders of magnitude smaller than the measured value. 
Since the Shklovskii term scales linearly with distance, adopting the  NE2001 distance ($d_{\rm NE2001}\simeq 2.13$ kpc) increases the kinematic correction ($\simeq5.8\times10^{-14}$), but it is still tiny compared to the measured $\dot{P}_b$.  
The large excess orbital-period derivative therefore indicates the presence of an additional, as-yet unmodeled contribution. We discuss the implications of this orbital evolution in Section~\ref{subsec:Orbital characteristics}.

\subsection{Wide-band timing} \label{sec:wideband-timing}

\begin{figure*}[ht!]
\begin{center}
\includegraphics[width=\textwidth,angle=0]{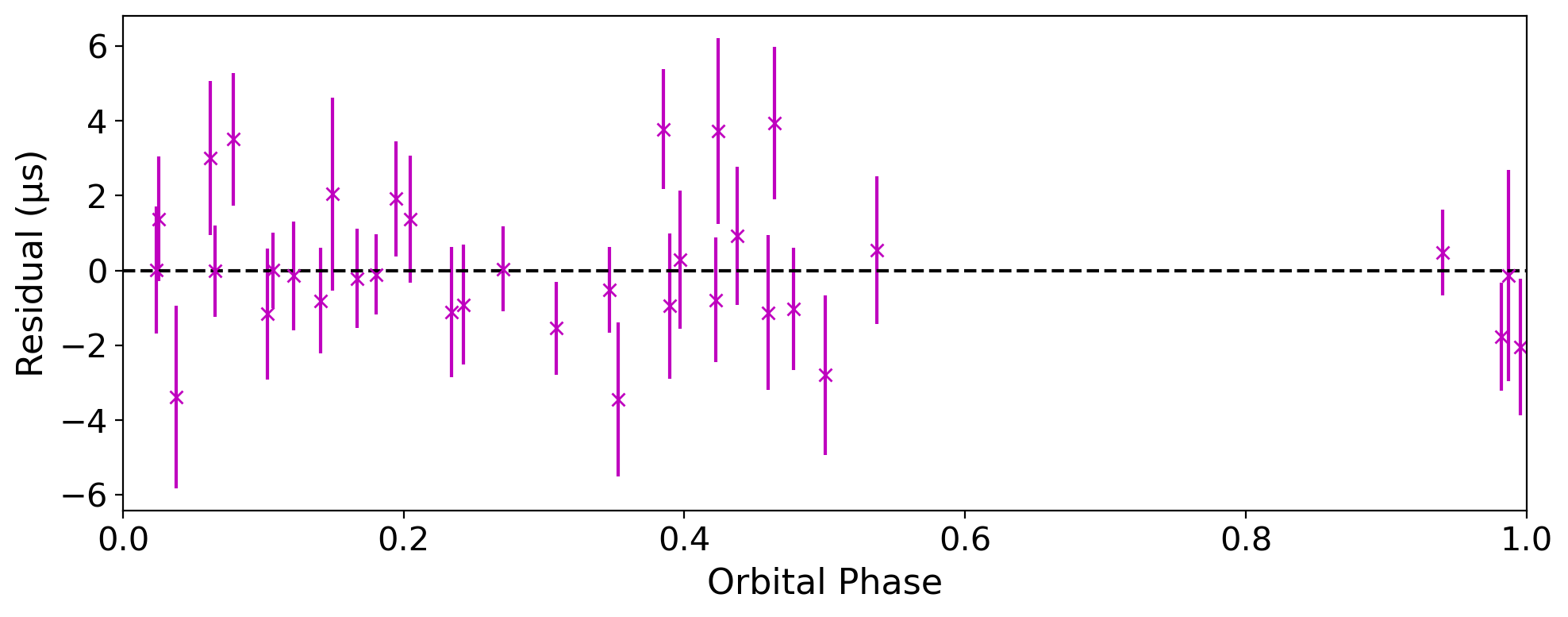}
\caption{Post-fit timing residual as a function of orbital phase from wideband timing of the Parkes UWL data only (our Parkes observations do not cover orbital phases 0.8--0.9.)}
\label{fig:WBtiming}
\end{center}
\end{figure*}

In addition to the conventional narrowband analysis, we also applied wide-band timing analysis to our Parkes UWL data using {\texttt{PulsePortraiture}} \citep {2019ApJ...871...34P} to account for the evolution of the pulsar profile with frequency while fitting for the ToAs and DM simultaneously. Using wide-band timing, we were able to reach a timing precision of 1.5 $\mu$s (Figure \ref{fig:WBtiming}) and also obtained DM residuals 0.0015~pc~cm$^{-3}$ (Figure \ref{fig:flux-pol}). For GMRT observations, we have not observed a significant improvement in wide-band timing compared to narrowband.

\subsection{Profile evolution with frequency}\label{sec:Profile evolution}

We first derived a phase-connected timing solution using only the uGMRT band 3 observations, and then used this solution to fold all subsequent observations from the other receivers, ensuring a consistent pulse-phase reference across backends. In the final timing analysis, we introduced JUMP parameters between each receiver/backend combination to absorb any remaining constant phase offsets due to instrumental delays or profile-evolution effects.

\begin{figure}[ht!]
\begin{center}
\includegraphics[width=0.4\textwidth,angle=0]{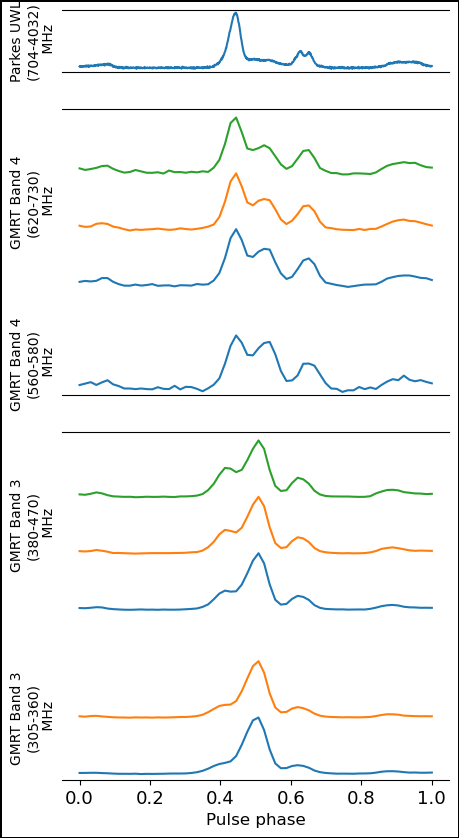}
\caption{Normalized, averaged pulse profiles from uGMRT observation at 400~MHz and 650~MHz, and Parkes UWL observation at 2368~MHz (bottom to top). The pulse profiles in this figure are aligned using the initial uGMRT Band 3 timing solution as the common reference. Different color in the plot shows different frequency ranges within an observing band. }
\label{fig:prof-evol}
\end{center}
\end{figure}

The pulse profile exhibits significant morphological evolution with frequency (Figure \ref{fig:prof-evol}), revealing changes not only in the overall width and scattering but also in the relative prominence of individual components. At the lowest frequencies from 305 to 360~MHz, the profile is dominated by a broad central component, with a hint of trailing emission on the right and no significant leading component on the left. As frequency increases to 380--470~MHz, the profile sharpens, and a secondary peak begins to emerge on the left side of the main pulse. This left component becomes increasingly prominent in uGMRT band 4 (550--750~MHz), eventually overtaking the original central peak to become the new dominant component. In parallel, the main peak at lower frequencies (observed at 305--360~MHz) gradually decreases in relative amplitude and becomes less distinct at higher frequencies. At the frequencies of Parkes UWL (704--4032~MHz), this formerly dominant central peak has diminished significantly, and the leading peak clearly defines the pulse profile.
The right-hand component, visible as a trailing feature at low frequencies, also evolves with increasing frequency. It is broad and blended with the main pulse at the lowest bands, but becomes more distinct and separable in the higher frequency part of band 3 (380--470~MHz) and especially in band 4. At UWL frequencies, it becomes more detached from the main peak. The main component has a width at 50\% of the peak intensity ($W_{50}$) of approximately 0.10 $\pm$ 0.02~ms at a frequency of 2368~MHz and 0.17$\pm$0.04~ms at 400~MHz. 
The complex, multi-component profiles seen in many MSPs can reflect emission arising from more than one magnetospheric region, with some components produced on open field lines above the polar cap and others associated with the current sheet beyond the light cylinder \citep{2025arXiv251005778K}. In this picture, the strong frequency dependence in Figure~\ref{fig:prof-evol}, including the changing dominance of components with observing band, may indicate that different profile components have distinct spectral behaviour and potentially distinct emission sites.
This overall profile evolution suggests complex spectral behavior across the pulse components, potentially driven by differing spectral indices, intrinsic emission geometry, and propagation effects such as scattering and birefringence in the magnetosphere or the interstellar medium. 

\subsection{Polarization and orbital variability of PSR~J2101$-$4802}
\label{subsec:polarization}

\begin{figure*}[ht!]
\begin{center}
\includegraphics[width=0.8\textwidth,angle=0]{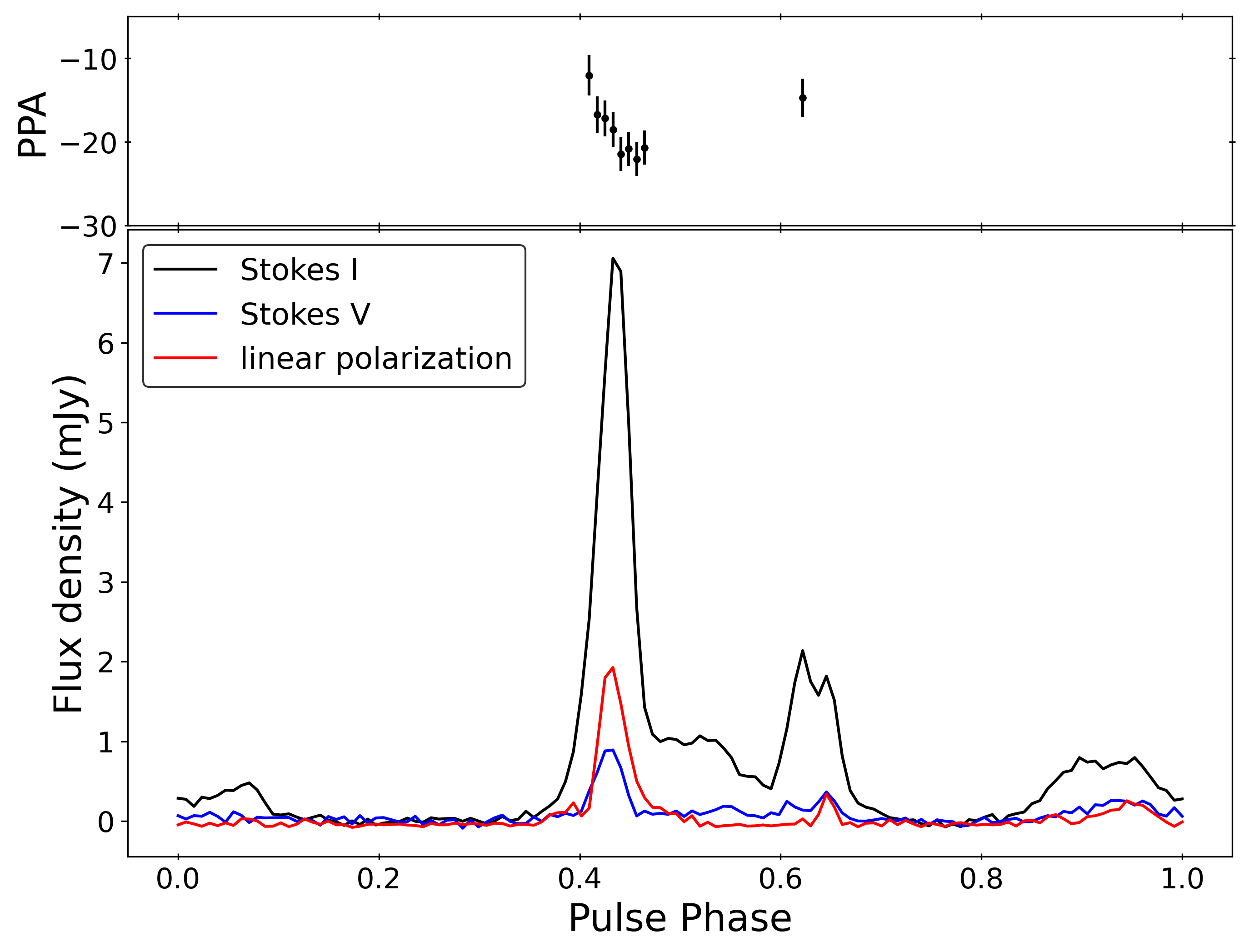}
\caption{Polarization profile at 2368 MHz from Parkes UWL observations. The upper panel shows the PPA variations (in black dots)}
\label{fig:pol}
\end{center}
\end{figure*}

\begin{figure*}[ht!]
\begin{center}
\includegraphics[width=0.8\textwidth,angle=0]{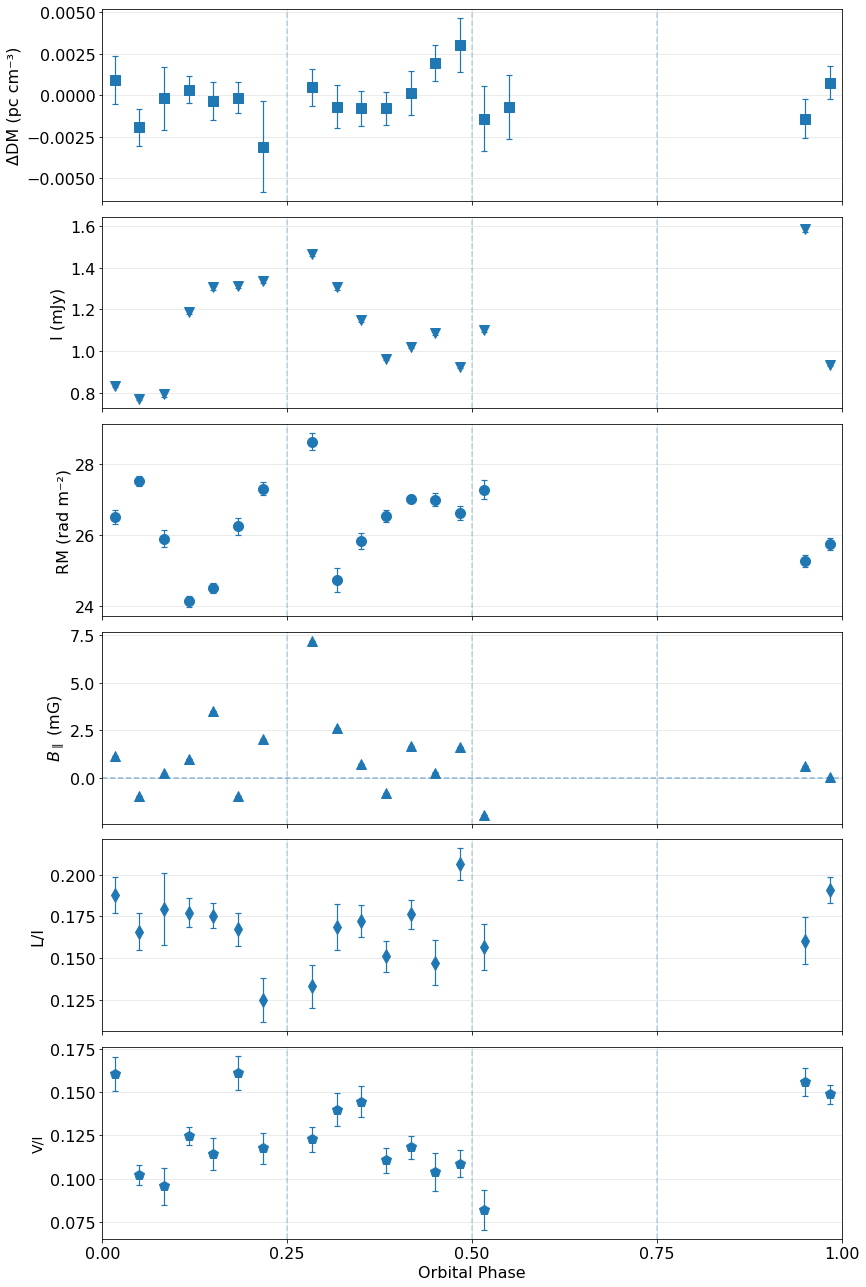}
\caption{Variation of DM, total intensity, RM, and parallel component of magnetic field along with polarization fractions as a function of orbital phase at 2368 MHz from the Parkes UWL observations.}
\label{fig:flux-pol}
\end{center}
\end{figure*}

 We analyzed Parkes UWL full-Stokes observations of
PSR~J2101$-$4802 spanning 708--4032~MHz. Following full
polarimetric calibration and rotation measure correction,
the orbit-averaged polarization profile
(Figure~\ref{fig:pol}) yields
$\mathrm{RM}=25.93\pm0.11~\mathrm{rad\,m^{-2}}$, with
mean linear and circular polarization fractions of
approximately 17\% and 12\%, respectively.  Flux densities
were estimated following
\citet{https://doi.org/10.3847/1538-4357/ac19b9} using
the full 3328~MHz bandwidth.  Orbital variation of
$\Delta\mathrm{DM}$ was obtained through wide-band timing
techniques (Section~\ref{sec:Timing}), along with
examination of the orbital-phase dependence of total
intensity, polarization properties, $\Delta\mathrm{DM}$,
and $\Delta\mathrm{RM}$ (Figure~\ref{fig:flux-pol}).
Within the frequency range of 708--2000~MHz, we examined
the polarization fractions as a function of frequency and
found no significant variation in either the linear or
circular polarization fraction across this band.

PSR~J2101$-$4802 exhibits no evidence for eclipses at
any orbital phase.  Nevertheless, we observe significant
phase-locked variability in multiple observables:
(i)~total intensity brightens toward $\phi\!\sim\!0.25$,
then declines between $\phi\!\approx\!0.4$--$0.55$ before
recovering; (ii)~$\Delta\mathrm{DM}$ displays weak overall
variability with a modest increase near $\phi\!\approx\!0.5$;
(iii)~RM exhibits modulations with orbital phase with peaks
at $\phi\!\approx\!0.05$, $\phi\!\approx\!0.25$, and
$\phi\!\approx\!0.5$; (iv) the linear polarization
fraction $L/I$ undergoes a deep suppression near
$\phi\!\approx\!0.25$ before recovering, while the
circular fraction $V/I$ reaches a general minimum near
$\phi\!\approx\!0.5$ though with significant scatter across all orbital phases.

$\mathrm{DM}=\int n_{e}\,dl$ traces the column density
of free electrons along the line of sight, while
$\mathrm{RM}=0.81\int n_{e} B_{\parallel}\,dl$ measures
the electron-weighted line-of-sight magnetic field of
the intervening plasma. The RM enhancements accompanied by only modest DM increases suggest localized regions with enhanced magnetic fields at multiple orbital phases dominate over the electron density variation along the line of sight, indicating a structured, non-uniform intrabinary medium.

Following \citet{2023ApJ...955...36W}, the implied
line-of-sight magnetic field strength of the intrabinary
plasma can be estimated by
\begin{equation}
    B_{\parallel}\;[\mu\mathrm{G}] \;\simeq\; 1.232\,
    \frac{\Delta\mathrm{RM}\;[\mathrm{rad\,m^{-2}}]}
         {\Delta\mathrm{DM}\;[\mathrm{pc\,cm^{-3}}]}\,,
    \label{eq:Bpar}
\end{equation}
where $\Delta\mathrm{RM}$ and $\Delta\mathrm{DM}$ denote
the orbital-phase-dependent variations, isolating the
intrabinary plasma contribution from the approximately
constant ISM foreground.  Applying
Equation~(\ref{eq:Bpar}) to the phase-resolved
measurements in Figure~\ref{fig:flux-pol} yields
characteristic magnetic fields at the milligauss level,
with peak values occurring near $\phi\!\approx\!0.25$.

 When the field is predominantly perpendicular to the line of sight, the natural modes become elliptical or linear, and generalised Faraday conversion refers to the exchange of power between $L/I$ and $V/I$ \citep{2019MNRAS.490..889P}; depending on the field configuration, this can either increase or decrease $V/I$ while simultaneously modifying $L/I$. The $V/I$ minima near $\varphi \approx 0.05$ and $\varphi \approx 0.5$, where $V/I$ reaches its lowest values while $L/I$ remains comparatively less suppressed, suggest that Faraday conversion preferentially transfers power from circular to linear polarization at these phases, consistent with a predominantly perpendicular field geometry. The elevated $V/I$ near $\varphi \approx 0.25$ further illustrates that conversion can also enhance $V/I$ depending on the local field configuration.

The $L/I$ suppression near $\varphi \approx 0.25$,
coinciding with the strongest RM enhancement but no
corresponding $V/I$ minimum, can be explained with a
predominantly parallel magnetic field geometry at that orbital
phase within the generalised Faraday rotation framework \citep{1998PASA...15..211K, 2010MNRAS.403..569W,
2019MNRAS.490..889P}. When
the intrabinary field is directed predominantly parallel
to the line of sight, the natural propagation modes are
circular, and normal Faraday rotation acts on the $L/I$ while leaving $V/I$ unaffected. Within-channel bandwidth depolarization gives a
differential PA rotation of
\begin{equation}
\begin{split}
    \delta\psi
        &= \mathrm{RM} \times \frac{2c^2\,\delta f}{f^3}
        \approx 26 \times
          \frac{2\times(3\times10^8)^2\times10^6}
               {(2.368\times10^9)^3} \\
        &\approx 3.5\times10^{-4}~\mathrm{rad}
        \approx 0.02\degr,
\end{split}
    \label{eq:bw_depol}
\end{equation}
corresponding to a depolarization factor of
$\approx\!0.99$, which is negligible.
Beam depolarization from spatial RM fluctuations
within the intrabinary medium \citep{2018ApJ...867...22Y},
\begin{equation}
    p_\mathrm{obs}
        = p_\mathrm{int}
          \times e^{-2\sigma_\mathrm{RM}^2\,\lambda^4},
    \label{eq:burn}
\end{equation}
where $\sigma_\mathrm{RM}$ is the RM dispersion from
small-scale spatial structure and $\lambda = c/f$ is the
observing wavelength, would require
$\sigma_\mathrm{RM} \approx 25$~rad~m$^{-2}$ to reproduce
the observed drop in $L/I$ from $\sim\!0.20$ to
$\sim\!0.125$.  This is comparable to the mean RM itself
($\approx\!26$~rad~m$^{-2}$) and far exceeds the observed
RM variation of $\sim\!4$~rad~m$^{-2}$ (Figure~\ref{fig:flux-pol}). The $L/I$ suppression near $\varphi \approx 0.25$,
however, cannot be explained only by propagation effects at
our observing frequency of 2368~MHz.
The intrabinary magnetic field
at this phase probes a different strength and orientation
than at $\phi\!\approx\!0.5$, as similarly observed in
PSR~B1957$+$20 \citep{2019MNRAS.484.5723L} and
PSR~J2051$-$0827 \citep{2023ApJ...955...36W}. We therefore conclude that the $L/I$ variation near
$\varphi \approx 0.25$ for MSP J2101--4802 is mostly intrinsic with a small contribution from the propagation effects.

In summary, PSR~J2101$-$4802 demonstrates modest DM variability but exhibits pronounced, orbital-phase-dependent RM modulation and polarization changes without detectable eclipses. The large RM variations with comparatively small DM changes, indicate that line-of-sight magnetic-field fluctuations, rather than density variations, dominate the observed behavior. These signatures are most naturally explained by propagation through structured intrabinary plasma—plausibly a shocked, turbulent companion wind—featuring strong magnetic fields and small-scale spatial structure.

\subsection{Constraining the emission geometry} \label{sec:emission geometry}

\begin{figure}[ht!]
\begin{center}
\includegraphics[width=0.5\textwidth,angle=0]{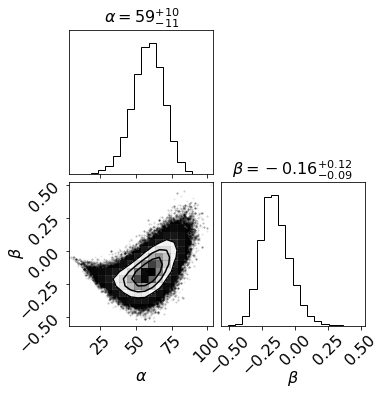}
\caption{Corner plot showing the posterior distributions from fitting the RVM to the position angle variation observed by the Parkes UWL.}
\label{fig:rvm}
\end{center}
\end{figure}

The polarization position angle (PPA) across the main pulse of PSR~J2101--4802 shows a hint of a swing over the pulse profile. We modelled this behaviour with the Rotating Vector Model (RVM; Equation~\ref{rvm}, \citealt{1969ApL.....3..225R}) to constrain the emission geometry:
\begin{equation}
\label{rvm}
\tan(\psi - \psi_0) = \frac{\sin\alpha \sin(\phi - \phi_0)}{\sin(\alpha + \beta)\cos\alpha - \cos(\alpha + \beta)\sin\alpha \cos(\phi - \phi_0)}.
\end{equation}
Here, $\phi$ denotes the pulsar rotational phase, while $\alpha$ and $\beta$ are the magnetic latitude and sightline impact angle, respectively. The parameters $\psi_0$ and $\phi_0$ represent the PPA and phase offsets. Since $\psi_0$ corresponds to the position angle at the point of steepest gradient in the PPA swing, we estimated $\psi_0$ and $\phi_0$ by maximizing the local rate of change of the PPA. This procedure yielded $\psi_0 = 34^{\circ}$ and $\phi_0 = 176^{\circ}$. We then used a Markov Chain Monte Carlo (MCMC) analysis to explore the posterior distributions of $\alpha$ and $\beta$. The median values with $1\sigma$ uncertainties are $\alpha = 59^{+10^{\circ}}_{-11^{\circ}}$ and $\beta = -0.16^{+0.12^{\circ}}_{-0.10^{\circ}}$. The banana-shaped joint posterior distribution (Figure~\ref{fig:rvm}) highlights the intrinsic degeneracy between these angles in the fitted model.

Previous polarimetric studies of MSPs have shown that, although their emission shares many properties with that of normal pulsars, their PPA swings only rarely conform to the simple RVM. Multifrequency studies by \citet{1999ApJ...526..957K} demonstrated that MSP emission can be broadly understood within an overall dipolar geometry, but exhibits substantial depolarization and profile complexity at high radio frequencies. Building on this, \citet{2011MNRAS.414.2087Y} presented 1.4-GHz polarization profiles for 20 Parkes Pulsar Timing Array MSPs and found that, in most cases, the PPA variations are inconsistent with the RVM, with only three MSPs (PSRs J1022+1001, J1744$-$1134, and J2124$-$3358) showing reasonably smooth, RVM-like swings and several others exhibiting essentially flat PPA swings. A polarization study of 24 MSPs by \citet{2015MNRAS.449.3223D} showed that, for almost all of their sample, the PPA swings across the often very wide pulse profiles cannot be fitted with a single RVM, with~PSR J1022+1001 at 10~cm being the only clear case of a near-classical S-shaped RVM swing. Taken together, these studies imply that very few MSPs exhibit global RVM-like PPA swing. Additionally, this behaviour can be frequency dependent, suggesting emission over extended regions high in the magnetosphere, with strong distortions arising from the superposition of orthogonal modes, magnetospheric propagation, and multi-altitude or patchy emission zones superposed on an underlying dipolar field \citep{1999ApJ...526..957K,2011MNRAS.414.2087Y,2015MNRAS.449.3223D}.

\section{Discussion}\label{sec:Discussion}

\subsection{Solar-wind contribution to timing residuals}

We obtained the excess DM from timing and computed the pulsar--Sun--observer angle $\rho(t)$ at each DMX epoch. The standard pulsar timing approach constrains the DM contribution from the solar wind (SW), approximating the SW as a spherically symmetrical distribution of electrons $n_{\mathrm{e,sw}}$ \citep{2006MNRAS.372.1549E}:
\begin{equation}
    n_{\mathrm{e,sw}} = A_{\mathrm{AU}}
    \left[ \frac{1~\mathrm{AU}}{r} \right]^2,
    \label{eq:ne_sw}
\end{equation}
where $r$ is the distance between the pulsar and the Sun and $A_{\mathrm{AU}}$ is the free electron density of the SW at 1~AU. The DM contribution of this model is obtained by integrating Equation~\ref{eq:ne_sw} along the line-of-sight (LoS), and it can be expressed as  \citep{2006MNRAS.372.1549E, 2021A&A...647A..84T}:
\begin{equation}
    \mathrm{DM}_{\mathrm{sw}} = 4.85 \times 10^{-6} \,
    A_{\mathrm{AU}} \, \frac{\rho}{\sin \rho} \,
    \mathrm{pc~cm^{-3}},
    \label{eq:DM_sw}
\end{equation}

\begin{figure*}
  \centering
  \includegraphics[width=\textwidth]{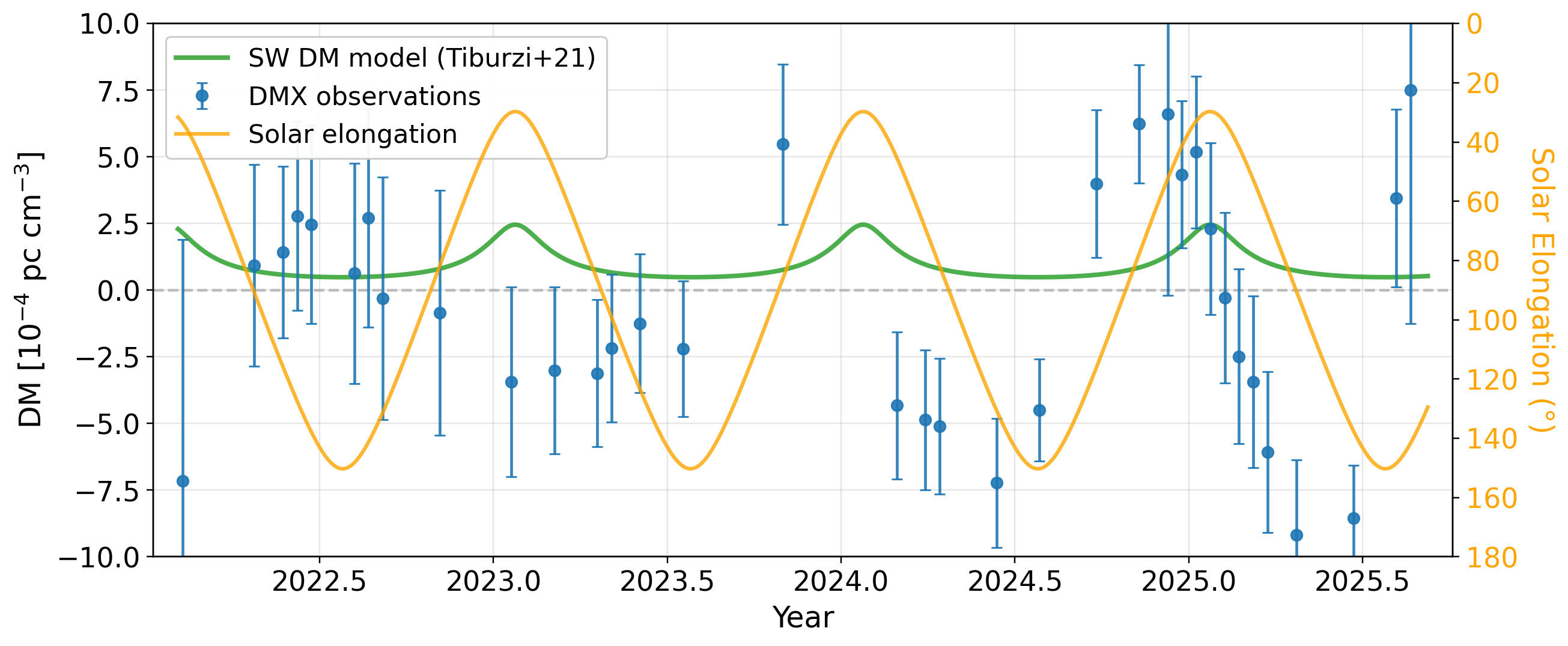}
  \caption{DMX time series (blue), best-fitting symmetric solar-wind model (green), and solar elongation (orange; right axis, inverted). The model follows the annual enhancement near small elongations and under-predicts the largest excursions, pointing to non-spherical solar-wind structure.}
  \label{fig:dmx_sw}
\end{figure*}

Figure~\ref{fig:dmx_sw} shows the solar wind contribution to the observed excess DM. The blue points are the measured DMX values (left axis, units of $10^{-4}\,\mathrm{pc\,cm^{-3}}$), the green curve is the best-fitting symmetric solar-wind model sampled at the DMX epochs, and the orange line (right axis, inverted) is the solar elongation. The short-term enhancements in DM at small solar elongations are suggestive of transient or asymmetric solar-wind structures, such as coronal mass ejections (CMEs), which are known to introduce localized increases in electron density along the line of sight \citep[e.g.,][]{2007MNRAS.378..493Y, 2021A&A...647A..84T, 2019ApJ...872..150M}. However, the observed DM variations cannot be fully explained by the symmetric solar wind model alone, as can be seen in Figure~\ref{fig:dmx_sw}. The residual variation in DM likely arises from a combination of structured solar wind features, turbulent ISM structures along the line of sight, and contributions from the intrabinary medium, as evidenced by the orbital-phase-locked DM variability discussed in Section~\ref{subsec:polarization}.

\subsection{Orbital characteristics}\label{subsec:Orbital characteristics}

Binary evolution models \citep[e.g.,][]{2014ApJ...786L...7B} predict that MSPs with He–WD companions end the spin-up phase with a low-mass, degenerate donor that underfills its Roche lobe \citep{2002ApJ...565.1107P}. Such systems generally lack strong irradiation/tidal optical modulations and rarely show radio eclipses. Our target, PSR~J2101$-$4802, having a median companion mass of $\sim0.15\,M_\odot$, an orbital period of $\sim1$~day, shows no eclipses, and has no optical counterpart, consistent with a detached He--WD companion. The measured spin-down power, $\dot{E}\sim10^{32}\ \mathrm{erg\ s^{-1}}$ is also typical of Galactic He--WD MSPs.

As discussed by \citet{2024ApJ...965...64G}, within the known population, nearly 20 He–WD MSP binaries with comparable $P_b$ and $M_c$ are listed in the ATNF catalogue \citep{2005AJ....129.1993M}; only six have measured $\dot{P}_b$; three Galactic field systems: PSRs J0751$+$1807, J1012$+$5307, J1738$+$0333; and three in globular clusters: PSRs J0024$-$7204U/Y and J1701$-$3006B. Galactic field systems shows $\dot{P}_b$ around $\sim10^{-14}$, whereas cluster systems show $\dot{P}_b\sim10^{-12}$ because of additional globular-cluster accelerations. In this context, PSR~J2101$-$4802, being a galactic field MSP, shows substantial orbital-period variability ($\dot{P}_b\sim10^{-11}$) unlike other Galactic He$-$WD MSPs, which cannot be explained by usual kinematic contributions to the $\dot{P}_b$.\par

Binary evolution models that include irradiation feedback and pulsar-driven ablation \citep[e.g.,][]{2014ApJ...786L...7B} predict that systems emerging from the low-mass X-ray binary (LMXB) phase may pass through an interacting redback-like stage before settling into detached MSP$-$He-WD binaries. During this phase, cyclic mass transfer and evaporation can strip the donor, leaving behind residual circumstellar material and a low-mass companion that eventually becomes a He--WD after extensive stripping. In spider redback systems, large and often variable orbital period derivatives ($\dot{P}_b \sim 10^{-12}$–$10^{-10}$) are commonly observed and are attributed to angular momentum exchange driven by mass loss, irradiation-induced winds, or changes in the companion’s quadrupole moment \citep[e.g.,][]{1992ApJ...385..621A,2011MNRAS.414.3134L,2015ApJ...807...18P,2013IAUS..291..127R}. PSR~J2101$-$4802 exhibits properties consistent with a detached He--WD system, including $M_c\sim0.15\,M_\odot$, $P_b\sim1$~day, the absence of radio eclipses, and a typical spin-down power; however, it shows an anomalously large orbital-period derivative compared to other Galactic He--WD MSPs. If PSR~J2101$-$4802 has only recently evolved out of an interacting phase, residual mass loss of the companion or torque by the circumbinary material on the companion still produces enhanced orbital evolution, yielding a $\dot{P}_b$ larger than typical He--WD systems but lower than classical spiders. The combination of orbital period, companion mass, spin-down power, and relatively higher $\dot{P}_b$ (Section~\ref{sec:Timing}), along with the signatures of magnetized plasma structures in the intrabinary material (Section~\ref{subsec:polarization}), suggests that PSR~J2101$-$4802 represents a transitional system between redback-like spiders and detached He--WD MSP binaries. This is one of the rare observational evidence connecting the formation of redbacks to He--WD binaries.\par

 While PSR~J2101--4802 lies broadly in the He--WD MSP region, its ($P_b$, $M_c$) location may not fall exactly on the canonical $P_b$--$M_{\rm WD}$ relation if one adopts the median inclination estimate $i=60^\circ$ ($M_c\simeq0.15\,M_\odot$). \citet{1999A&A...350..928T} derived an analytic $P_b$--$M_{\rm WD}$ relation for detached LMXB remnants, which, when extrapolated to $P_b \simeq 1$~day, predicts a somewhat larger He--WD mass of $\sim 0.18-0.20\,M_\odot$. Although \citet{1999A&A...350..928T} calibrated their relation primarily for $P_b \gtrsim 2$~days, \citet{2014A&A...571A..45I} demonstrated that the relation remains broadly valid at shorter orbital periods, albeit with increased scatter in the $M_{\rm WD}\sim0.15-0.20\,M_\odot$ range due to details of detachment process and residual envelope stripping. Using the measured mass function, agreement with a predicted companion mass of $M_c \sim 0.18-0.20\,M_\odot$ would require a more modest inclination of $i \simeq 42^\circ-48^\circ$ (for $M_p=1.4\,M_\odot$). In this context, other systems with non-standard or “transitional” companions--such as PSR~J1816+4510, whose companion is hot, inflated, and characterized by unusually low surface gravity \citep{2013ApJ...765..158K, 2014ApJ...780..167K}--illustrate that some MSP binaries may occupy intermediate evolutionary states between interacting redback-like systems and fully detached He–WD MSPs.\par

\subsubsection{Duration of the transitional phase and connection to $\ddot{P}_b$} \label{sec:transition_duration}

A physical estimate of the duration of the transitional phase can be made from the proto-He~WD contraction timescale $\Delta t_\mathrm{proto}$, i.e., the time from Roche-lobe detachment until the companion settles onto the WD cooling track. Throughout this phase, the companion
remains bloated and burns residual hydrogen in its envelope \citep{2014A&A...571L...3I, 2016A&A...595A..35I}. In this stage, the system is capable of sustaining enhanced orbital variability through tidal coupling or quadrupole moment variations. After settling this phase, $\dot{P}_b$ decays toward the canonical kinematic value \citep{2009MNRAS.400..805L}.\par
\citet{2014A&A...571L...3I} compute $\Delta t_\mathrm{proto}$ from the stellar evolution models incorporating mass transfer, gravitational radiation, magnetic braking, and residual hydrogen shell burning. The Figure~3 of \citet{2014A&A...571L...3I} shows that $\Delta t_\mathrm{proto}$ spans 96--1910~Myr for companion masses of 0.24--0.17~$M_\odot$ respectively. The 96~Myr end corresponds to the highest mass (with hydrogen shell flashes) and the 1910~Myr end corresponds to the lowest mass (without flashes). This strong dependence on $M_\mathrm{WD}$ arises from the core mass--luminosity relation
$L \propto M_\mathrm{WD}^7$ \citep{1970A&A.....6..426R}. Lower-mass He--WDs are less luminous and therefore burn their residual hydrogen envelope (${\sim}\,0.01~M_\odot$) more slowly. The \citet[Equation~1 and Appendix~B of][]{2014A&A...571L...3I} gives
\begin{equation}
    \Delta t_\mathrm{proto}
        \approx 400~\mathrm{Myr}
          \times \left(\frac{0.20~M_\odot}{M_\mathrm{WD}}
                 \right)^{\!7},
    \label{eq:istrate}
\end{equation} accurate to within $\sim\!50\%$ for the no hydrogen shell flash mass range of He--WD.

\citet{2014A&A...571L...3I} find that hydrogen shell flashes mostly occur above $M_\mathrm{WD} \geq 0.21~M_\odot$. Above this threshold, the flashes erode the hydrogen envelope faster, resulting in a smaller value of $\Delta t_\mathrm{proto}$. Low-mass WDs ($< 0.20~M_\odot$) avoid flashes entirely and remain bloated for a long time. {\bf Since the orbital inclination of PSR~J2101$-$4802 is unconstrained, the companion mass ranges from $0.13~M_\odot$ at $i = 90\degr$ (minimum, edge-on) to $0.15~M_\odot$ at $i = 60\degr$ (median) and $0.32~M_\odot$ at $i = 26\degr$ (maximum), as listed in Table~\ref{tab:2}.} Therefore, the minimum and median companion mass (0.13 and 0.15~$M_\odot$) for PSR~J2101--4802 lie well below the threshold for shell flash to occur. Table~\ref{tab:dtproto} lists $\Delta t_\mathrm{proto}$ calculated for different inclination angles (using Equation~\ref{eq:istrate}). We note from Figure~3 of \citet{2014A&A...571L...3I} that $\Delta t_\mathrm{proto}$ $\approx 1910$~Myr for a He--WD of mass 0.17~$M_\odot$. 
Thus, considering Equation~\ref{eq:istrate}, and minimum companion mass 0.13~$M_\odot$, $\Delta t_\mathrm{proto}$ for  PSR~J2101$-$4802 exceeds 1910~Myr.
\begin{table}[ht]
\centering
\caption{Proto-He--WD contraction timescale and predicted
         $\ddot{P}_b$ for the companion mass range of
         PSR~J2101$-$4802 corresponding to the orbital
         inclinations of $90\degr$, $60\degr$, and
         $18\degr$.
         $^*$above the shell-flash threshold.}
\label{tab:dtproto}
\begin{tabular}{l@{\hspace{8pt}}c@{\hspace{8pt}}c@{\hspace{8pt}}c@{\hspace{12pt}}c}
\hline\hline
Inclination
    & $M_c$~($M_\odot$)
    & Flashes
    & $\Delta t_\mathrm{proto}$
    & $\ddot{P}_b$~(s\,s$^{-2}$) \\
\hline
$i = 90\degr$ & 0.13 & No  & ${\sim}9.4$~Gyr          & $7.7\times10^{-29}$ \\
$i = 60\degr$ & 0.15 & No  & ${\sim}3.2$~Gyr          & $2.4\times10^{-28}$ \\
$i = 26\degr$ & 0.32 & Yes$^*$ & ${\sim}16$~Myr$^*$   & $2.4\times10^{-26*}$ \\
\hline
\end{tabular}
\end{table}

Once the WD companion cools down, intrinsic orbital variability decreases, and the observed
$\dot{P}_b$ becomes dominated entirely by kinematic contributions. For PSR~J2101$-$4802, we computed in Section~\ref{sec:kinematics} the Shklovskii and Galactic
acceleration contributions, $\dot{P}_b^\mathrm{final} \approx \dot{P}_b^\mathrm{D}
+ \dot{P}_b^\mathrm{GW} = (2.7 \pm 1.3)\times10^{-14}~\mathrm{s\,s^{-1}}$.
This is consistent with the kinematically dominated $\dot{P}_b$ values seen in analogous Galactic-field He--WD MSPs. For example, PSR~J0751$+$1807, PSR~J1012$+$5307, and PSR~J1738$+$0333, all have
$|\dot{P}_b^\mathrm{obs}| \sim 10^{-14}~\mathrm{s\,s^{-1}}$ \citep{2009MNRAS.400..805L, 2012MNRAS.423.3328F}.

Moreover, the non-detection of the companion in optical and near-infrared imaging (Section~\ref{sec:Multi-Wavelength}) implies that the companion is no longer in the early, luminous proto-He~WD phase but has already faded significantly. This suggests that the companion of PSR~J2101--4802 could be in the later stages of the proto-WD contraction phase and is approaching the WD cooling track.\par  
In absence of the information on present transitional state of the MSP companion, we adopt half the total $\Delta t_\mathrm{proto}$ from Equation~(\ref{eq:istrate}) as an estimate of the remaining transitional timescale: $\Delta t^r_\mathrm{proto} \approx 1.6$~Gyr for $i =60^\circ$ ($M_c = 0.15~M_\odot$, half of 3.2~Gyr) and $\Delta t^r_\mathrm{proto} \approx 5.0$~Gyr for
$i = 90^\circ$ ($M_c = 0.13~M_\odot$, half of 9.4~Gyr).

The average rate of change of $\dot{P}_b$ over the remaining transitional phase provides a order-of-magnitude prediction for $\ddot{P}_b$:
\begin{equation}
    \ddot{P}_b
        \sim \frac{\dot{P}_b^\mathrm{now}
                   - \dot{P}_b^\mathrm{final}}
                  {\Delta t^r_\mathrm{proto}}
        = \frac{1.21\times10^{-11}
                - 2.7\times10^{-14}}
               {\Delta t^r_\mathrm{proto}}.
    \label{eq:pbdd_transition}
\end{equation}
Substituting the values of $\Delta t^r_\mathrm{proto}$, the predicted $\ddot{P}_b$ ranges from $7.7\times10^{-29}$ to $2.4\times10^{-28}~\mathrm{s\,s^{-2}}$
for the low-mass WD companions with no shell-flash. These are listed in the final column of
Table~\ref{tab:dtproto}.\par

Following \citet{2023ApJ...942...87K} (their Section~3.1.3, Equations~4--6), the orbital phase predicted by the \textsc{BTX} model at time $t$ relative to the reference epoch $t_0$ is
\begin{equation}
    \phi_\mathrm{orb}(t)
        = \phi_0
          + n_b\,(t-t_0)
          + \frac{\dot{n}_b}{2!}\,(t-t_0)^{2}
          + \frac{\ddot{n}_b}{3!}\,(t-t_0)^{3}
          + \cdots,
    \label{eq:btx_phase}
\end{equation}
where $n_b = 1/P_b$ is the orbital frequency and $\dot{n}_b$, $\ddot{n}_b$ are its first and second time derivatives.  If no orbital period evolution is present (all higher derivatives are zero), this reduces to the simple Keplerian prediction $\phi_\mathrm{Kep}(t) = \phi_0 + n_b(t-t_0)$.
Subtracting the Keplerian prediction from the \textsc{BTX} phase gives the deviation in orbital phase,
\begin{equation}
    \Delta\phi(t)
        = \frac{\dot{n}_b}{2!}\,(t-t_0)^{2}
          + \frac{\ddot{n}_b}{3!}\,(t-t_0)^{3}
          + \cdots
    \label{eq:delta_phi}
\end{equation}
The $\Delta T_0$ corresponding to epoch of
ascending node and $\Delta\phi$ are related as \citep[Equation~6 of][]{2023ApJ...942...87K}
\begin{equation}
    \Delta T_0 = \Delta\phi \times P_b.
    \label{eq:delta_T0}
\end{equation}
For a timing baseline $T$, the dominant contribution from an unmodelled $\ddot{P}_b$ enters through the $\ddot{n}_b$ term.  Using $\ddot{n}_b \approx -\ddot{P}_b/P_b^2$ to leading order, the amplitude of the resulting timing residual is
\begin{equation}
    \Delta T_0
        \approx P_b \cdot \frac{|\ddot{n}_b|}{3!}\,T^3
        = \frac{|\ddot{P}_b|}{6\,P_b}\,T^{3}.
    \label{eq:timing_effect}
\end{equation}
Setting $\Delta T_0 = \sigma_\mathrm{rms}$, i.e.\ when the cubic signal grows large enough to equal the rms of the timing residuals, and solving for the baseline $T$ required for detection,
\begin{equation}
    T_\mathrm{detect}
        = \left(\frac{6\,\sigma_\mathrm{rms}\,P_b}
                     {|\ddot{P}_b|}\right)^{\!1/3}.
    \label{eq:T_detect}
\end{equation}
Adopting the predicted $\ddot{P}_b \sim
7.7\times10^{-29}$--$2.4\times10^{-28}$~s\,s$^{-2}$, $\sigma_\mathrm{rms} = 3.50\times10^{-6}$~s, and
$P_b = 0.99$ days,
\begin{align}
    T_\mathrm{detect} 
         &\approx 60-90~\mathrm{yr}
    \label{eq:T_detect_values}
\end{align}
The predicted $\ddot{P}_b \sim 10^{-29}$--$10^{-28}$~s\,s$^{-2}$ from the proto-He--WD contraction scenario would therefore require a timing baseline of $\sim\!60$--$90$~yrs, to become detectable at the current timing precision.\par

Moreover, Equation~(\ref{eq:T_detect}) also allows us to obtain a minimum value of $\ddot{P}_b$ from the 3.7-yr dataset. Considering the timing noise of $\sigma_\mathrm{rms} = 3.50~\mu$s, we get,
\begin{equation}
\begin{split}
    \ddot{P}_b^\mathrm{min}
        &= \frac{6\,\sigma_\mathrm{rms}\,P_b}{T^3} \\
        &= \frac{6\times3.50\times10^{-6}
                 \times 86334.18}
                {(1.168\times10^8)^3} \\
        &\approx 1.1\times10^{-24}~\mathrm{s\,s}^{-2}.
\end{split}
    \label{eq:pbdd_sensitivity}
\end{equation}
Any $\ddot{P}_b$ above this level would, in principle, produce a detectable trend in the timing residuals.


\subsection{Search of multi-wavelength counterparts}\label{sec:Multi-Wavelength}

\subsubsection{X-ray and optical}\label{sec:X-ray and optical}

\begin{figure*}[htbp]
\includegraphics[width=\textwidth]{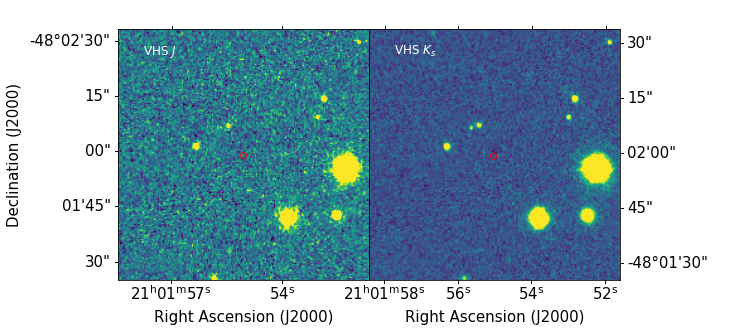}
\caption{VHS $J$ (1.2\,$\mu$m, left) and $K_s$ (2.1\,$\mu$m, right) near-infrared  images of the field of PSR J2101$-$4802.  The position of PSR J2101$-$4802 is marked with a red circle.}
\label{fig:image}
\end{figure*}

We have searched for X-ray counterparts using the radio timing position in existing \textit{Chandra}, \textit{XMM-Newton}, or \textit{Neil Gehrels Swift Observatory} archival data\footnote{\url{https://cxcfps.cfa.harvard.edu/cda/footprint/cdaview.html}}. No X-ray observation is available. 

We also searched for an X-ray counterpart in the \textit{eROSITA} all-sky survey. No eRASS1/DR1 catalog source \citep{2024A&A...682A..34M} is found within a $1\arcmin$ radius of the radio timing position. Using the eROSITA DR1 flux upper-limit  at the pulsar timing position in the $0.2$--$2.3$~keV band (band~024), we derive a $3\sigma$ flux upper limit of $F_X < 8.09\times10^{-14}$~erg~s$^{-1}$~cm$^{-2}$ (\texttt{Flag\_pos}=0 and $\mathrm{UL_B}=\mathrm{UL_S}$; \citealt{2024A&A...682A..35T}).

To estimate an unabsorbed luminosity limit, we correct for interstellar absorption by adopting $N_{\rm H}=7.5\times10^{20}$~cm$^{-2}$ inferred from the DM--$N_{\rm H}$ relation of \citet{2013ApJ...768...64H}.
Assuming a power-law spectrum with photon index $\Gamma=1.7$, WebPIMMS{\footnote{\url{https://heasarc.gsfc.nasa.gov/cgi-bin/Tools/w3pimms/w3pimms.pl}}} yields an absorption-corrected (unabsorbed) flux limit of
$F_{X,{\rm unabs}} < 1.15\times10^{-13}$~erg~s$^{-1}$~cm$^{-2}$ in the same band.
For $d= 1$~kpc, this corresponds to an unabsorbed luminosity limit of
$L_{X,{\rm unabs}} < 4\pi d^2 F_{X,{\rm unabs}} \simeq 1.35\times10^{31}$~erg~s$^{-1}$.
For comparison, rotation-powered pulsars (including MSPs) typically exhibit $L_X \sim 10^{-3}\dot{E}$ (with substantial scatter; \citealt{1997A&A...326..682B,2009ASSL..357...91B}), which for PSR~J2101$-$4802 ($\dot{E}\simeq 6.0\times10^{32}$~erg~s$^{-1}$) implies an expected $L_X \sim 6\times10^{29}$~erg~s$^{-1}$, i.e. well below our eROSITA limit and therefore consistent with the non-detection.

We searched for available optical/near-infrared surveys to constrain the counterpart to PSR~J2101$-$4802.  The deepest datasets we found were SkyMapper \citep{2018PASA...35...10W,2019PASA...36...33O} in the optical, and the Vista Hemisphere Survey (VHS; \citealt{2013Msngr.154...35M}) in the near-infrared.  We show the $J$ (1.2\,$\mu$m) and $K_s$ ($2.1\,\mu$m) images from VHS in Figure~\ref{fig:image}.  No source is visible within the sub-arcsecond error circle of PSR~J2101$-$4802 in any of the available data.  We set 5$\sigma$ limits of $g=20.1$, $r=20.6$, $i=21.3$ (AB) in the SkyMapper data based on the noise levels in nearby sources, and $J=20.0$ and $K_s=18.3$ (AB) in the VHS data based on the \texttt{ABMAGLIM} values in the data headers. We adopt a nominal distance of $d= 1$~kpc based on the pulsar's dispersion measure \citep{2002astro.ph..7156C,2017ApJ...835...29Y}. 
We estimate the line-of-sight extinction by first inferring the absorbing column from the dispersion measure using the empirical DM--$N_{\rm H}$ relation of \citet{2013ApJ...768...64H}, giving $N_{\rm H}\simeq 7.5\times10^{20}$~cm$^{-2}$. 
Using the $N_{\rm H}$--$A_V$ conversion of \citet{1995A&A...293..889P}, this corresponds to $A_V \simeq 0.42$~mag.
For $d= 1$~kpc, the distance modulus is $\mu = 5\log_{10}(d/10\,{\rm pc})\simeq 9.98$. We convert the apparent-magnitude limits to extinction-corrected absolute-magnitude limits via
$M_\lambda = m_\lambda - \mu - A_\lambda$ \citep[e.g.,][]{1998ApJ...500..525S},
where $\mu$ is the distance modulus and $A_\lambda$ is the extinction in band $\lambda$.
We adopt standard extinction ratios $A_g/A_V$, $A_r/A_V$, and $A_i/A_V$ from \citet{2018PASA...35...10W} and $A_J/A_V$ and $A_{K_s}/A_V$ from \citet{1989ApJ...345..245C}.
Adopting standard extinction ratios ($A_g\simeq 0.963A_V$, $A_r\simeq 0.738A_V$, $A_i\simeq 0.512A_V$, $A_J\simeq 0.282A_V$, $A_{K_s}\simeq 0.112A_V$), we found the extinction-corrected absolute-magnitude limits of
$M_g \gtrsim 9.6$, $M_r \gtrsim 10.3$, $M_i \gtrsim 11.1$, $M_J \gtrsim 10.3$, and $M_{K_s} \gtrsim 9.7$.
These extinction-corrected absolute-magnitude limits can be compared to He--WD cooling models appropriate for MSPs with low-mass companions. In binary-evolution calculations for He--WD remnants with masses $\simeq 0.16$--$0.20\,M_\odot$, the companion is expected to spend a ``proto-WD'' phase after detachment that can last $\sim10^8$--$10^9$~yr, during which it can be relatively luminous (bloated and hot), before settling onto the cooling track; the subsequent cooling rate depends sensitively on the residual H-envelope mass and the occurrence of shell flashes, leading to a wide range of optical luminosities at a given age \citep{2002MNRAS.337.1091S,1999MNRAS.303...30B,2016A&A...595A..35I,2016ApJ...830...36A}.
Therefore, at the distance/extinction of PSR~J2101--4802, our limits of $M_g\gtrsim 9.6$ and $M_r\gtrsim 10.3$ (AB) imply that we would have detected only the more luminous end of the expected companion distribution, whereas an $\sim0.15\,M_\odot$ He--WD can readily be fainter than these limits \citep[e.g.,][]{2016A&A...595A..35I,2006A&A...456..295B} as its luminosity can drop to levels corresponding to absolute magnitudes fainter than our observational limits. Thus, the current SkyMapper/VHS non-detections do rule out a very luminous companion at the pulsar position; substantially deeper optical imaging would be required to constrain the He--WD properties.\par

\subsubsection{Gamma-ray}\label{sec:Gamma-ray}
The only \textit{Fermi}-LAT 4FGL-DR4 GeV gamma-ray source near PSR~J2101$-$4802 is $1.^\circ 1$ away and is associated with a probable blazar \citep{4FGL-DR3, 4FGL-DR4,2009ApJ...697.1071A}. The forthcoming 16-yr LAT source list similarly reveals no TS\,$>$\,25 detection at the pulsar position. 

We used our radio timing ephemeris to calculate the phases of $17.4$~yr of LAT photons above 100~MeV and within $2^\circ$ of the pulsar, weighted as per \citet{SearchPulsation}. We use the weekly photon files available at \url{https://fermi.gsfc.nasa.gov/ssc/data/access/}, which are \texttt{SOURCE} class processed with Pass 8 r3 \citep{Pass8,improvedPass8}. Pulsation searches are generally insensitive to data selection details. No pulsed signal appears.

Nevertheless, the pulsed H-test test statistic rises steadily with MJD, reaching $H\!\sim\!17$--$20$ (i.e.\ $\sim\!3\sigma$ before accounting for trials).

The pulse profile is nearly sinusoidal, which is common among false positives and therefore warrants caution. 

Parameter files derived from radio observations provide highly reliable position, $F_0$, and orbital parameters. We tested sensitivity to the small spin-down by scanning $F_1$; a modest tweak from the timing value $F_1=-1.778\times10^{-16}\ \mathrm{s^{-2}}$ to $F_1=-1.8\times10^{-16}\ \mathrm{s^{-2}}$ improves the visual alignment of a string of near-vertical photons in the MJD–phase plane and yields the highest $H$, but the overall significance remains below the H-test $>25$ threshold established by \citet{ThousandFold}. The H-test evolution shows slope changes consistent with LAT exposure variations, and we find no evidence for variability in the pulsed or steady (non-pulsed) gamma-ray flux. We estimate an upper limit to the energy flux above 100~MeV near $6 \times 10^{-13 }$ erg cm$^{-2}$ s$^{-1}$, using the sensitivity map provided in Section 7.2 of the gamma-ray pulsar catalog \citep[][hereafter 3PC]{3PC}. At 1~kpc, the luminosity upper limit is then $< 7 \times 10^{-31 }$~erg~s$^{-1}$. The only MSP in 3PC with $\dot E < 1 \times 10^{33}$~erg~s$^{-1}$ has roughly that luminosity, favoring the idea that PSR~J2101$-$4802 may have a gamma-ray flux just below the limit of the LAT's sensitivity.

Given (i) the lack of a cataloged DC source, (ii) the marginal pulsation significance with a sinusoidal profile, and (iii) the trials incurred by exploring data partitions and $F_1$ refinements, we regard the present gamma-ray evidence as intriguing but inconclusive. If confirmed, PSR~J2101$-$4802 would extend the LAT MSP population slightly beyond the 3PC $(\dot{E},P)$ locus.

\section{Summary}\label{sec:Summary}

We report GMRT timing and orbital phase–resolved polarimetry of PSR~J2101−-4802, a 9.48-ms Galactic-disk MSP in a $\approx$1-day binary. Timing analysis indicates a likely MSP–He-WD system with a median companion mass $\approx$0.15 M$_{\odot}$ (assuming the inclination angle 60\degr), consistent with canonical recycling. Wideband full-Stokes data enable an RVM fit to the polarization position-angle swing, constraining the emission geometry (magnetic inclination and impact angle). The pronounced component-dependent profile evolution with frequency may indicate contributions from multiple emission regions, e.g., polar-cap emission plus a component associated with current-sheet emission beyond the light cylinder. This motivates caution in RVM fitting, since fitting only the polarization associated with likely polar-cap components may be more appropriate than using the full profile. The source’s precise localization and polarization properties make it a useful addition for testing MSP-HeWD binary evolution and probing magnetospheric geometry.
The timing solution shows an unusually large orbital period derivative, $\dot{P}_b$, relative to typical Galactic-field MSP-HeWD binaries. We evaluate Shklovskii and Galactic-acceleration terms plus GR damping and show that these kinematic/GR contributions can
not meet the measured $\dot{P}_b$. PSR~J2101$-$4802 thus provides rare observational evidence for an evolutionary pathway in which systems emerging from an interacting redback-like phase evolve toward detached He--WD MSP binaries, while still retaining residual orbital variability and magnetized-intrabinary structures. However, extending the timing baseline and more uniform orbital sampling at a broader range of local sidereal times will be critical to improve orbital phase coverage, reduce parameter covariances, and disentangle the intrinsic orbital evolution from timing systematics.

 \textit{Acknowledgements:} 

We acknowledge the support of the Department of Atomic Energy, Government of India, under project No.12-R\&D-TFR5.02-0700. DLK, LV, and GA acknowledge support
from the National Science Foundation (NSF) Physics Frontiers Center award numbers 1430284 and
2020265. AG acknowledges support from the Sarojini Damodaran Fellowship\footnote{\url{https://www.tifr.res.in/endowment/sdf/index.html}}, which enabled a research visit to the University of Wisconsin–Milwaukee (UWM) for collaboration on this work.
 We acknowledge the support of GMRT telescope operators for observations. The GMRT is run by the National Centre for Radio Astrophysics of the Tata Institute of Fundamental Research, India. The Parkes radio telescope is part of the Australia Telescope, which is funded by the Commonwealth of Australia for operation as a National Facility managed by CSIRO. We sincerely thank Mathew Bailes for valuable feedback and insightful suggestions to obtain the timing solution of this work.

The \textit{Fermi} LAT Collaboration acknowledges generous ongoing support
from a number of agencies and institutes that have supported both the
development and the operation of the LAT, as well as scientific data analysis.
These include the National Aeronautics and Space Administration and the
Department of Energy in the United States, the Commissariat \`a l'Energie Atomique
and the Centre National de la Recherche Scientifique / Institut National de Physique
Nucl\'eaire et de Physique des Particules in France, the Agenzia Spaziale Italiana
and the Istituto Nazionale di Fisica Nucleare in Italy, the Ministry of Education,
Culture, Sports, Science, and Technology (MEXT), High Energy Accelerator Research
Organization (KEK) and Japan Aerospace Exploration Agency (JAXA) in Japan, and
the K.~A.~Wallenberg Foundation, the Swedish Research Council, and the
Swedish National Space Board in Sweden. Additional support for science analysis during the operations phase is greatly 
acknowledged by the Istituto Nazionale di Astrofisica in Italy and the Centre 
National d'\'Etudes Spatiales in France. This work was performed in part under DOE 
Contract DE-AC02-76SF00515.

We thank the reviewer for the thoughtful comments and constructive suggestions, which have significantly improved the manuscript.

\bibliography{J2101}{}
\bibliographystyle{aasjournalv7}

\end{document}